\documentclass[12pt]{article}
\usepackage{epsfig}
\usepackage{amsmath}
\usepackage{hhline}
\usepackage{amssymb}
\usepackage{times}
\usepackage{cite}
\usepackage{longtable}
\usepackage{multirow}

\newlength{\dinwidth}
\newlength{\dinmargin}
\begin{document}  
\newcommand{\pom}{{I\!\!P}}
\newcommand{\reg}{{I\!\!R}}
\newcommand{\slowpi}{\pi_{\mathit{slow}}}
\newcommand{\fiidiii}{F_2^{D(3)}}
\newcommand{\fiidiiiarg}{\fiidiii\,(\beta,\,Q^2,\,x)}
\newcommand{\n}{1.19\pm 0.06 (stat.) \pm0.07 (syst.)}
\newcommand{\nz}{1.30\pm 0.08 (stat.)^{+0.08}_{-0.14} (syst.)}
\newcommand{\fiidiiiful}{F_2^{D(4)}\,(\beta,\,Q^2,\,x,\,t)}
\newcommand{\fiipom}{\tilde F_2^D}
\newcommand{\ALPHA}{1.10\pm0.03 (stat.) \pm0.04 (syst.)}
\newcommand{\ALPHAZ}{1.15\pm0.04 (stat.)^{+0.04}_{-0.07} (syst.)}
\newcommand{\fiipomarg}{\fiipom\,(\beta,\,Q^2)}
\newcommand{\pomflux}{f_{\pom / p}}
\newcommand{\nxpom}{1.19\pm 0.06 (stat.) \pm0.07 (syst.)}
\newcommand {\gapprox}
   {\raisebox{-0.7ex}{$\stackrel {\textstyle>}{\sim}$}}
\newcommand {\lapprox}
   {\raisebox{-0.7ex}{$\stackrel {\textstyle<}{\sim}$}}
\def\gsim{\,\lower.25ex\hbox{$\scriptstyle\sim$}\kern-1.30ex%
\raise 0.55ex\hbox{$\scriptstyle >$}\,}
\def\lsim{\,\lower.25ex\hbox{$\scriptstyle\sim$}\kern-1.30ex%
\raise 0.55ex\hbox{$\scriptstyle <$}\,}
\newcommand{\pomfluxarg}{f_{\pom / p}\,(x_\pom)}
\newcommand{\dsf}{\mbox{$F_2^{D(3)}$}}
\newcommand{\dsfva}{\mbox{$F_2^{D(3)}(\beta,Q^2,x_{I\!\!P})$}}
\newcommand{\dsfvb}{\mbox{$F_2^{D(3)}(\beta,Q^2,x)$}}
\newcommand{\dsfpom}{$F_2^{I\!\!P}$}
\newcommand{\gap}{\stackrel{>}{\sim}}
\newcommand{\lap}{\stackrel{<}{\sim}}
\newcommand{\fem}{$F_2^{em}$}
\newcommand{\tsnmp}{$\tilde{\sigma}_{NC}(e^{\mp})$}
\newcommand{\tsnm}{$\tilde{\sigma}_{NC}(e^-)$}
\newcommand{\tsnp}{$\tilde{\sigma}_{NC}(e^+)$}
\newcommand{\st}{$\star$}
\newcommand{\sst}{$\star \star$}
\newcommand{\ssst}{$\star \star \star$}
\newcommand{\sssst}{$\star \star \star \star$}
\newcommand{\tw}{\theta_W}
\newcommand{\sw}{\sin{\theta_W}}
\newcommand{\cw}{\cos{\theta_W}}
\newcommand{\sww}{\sin^2{\theta_W}}
\newcommand{\cww}{\cos^2{\theta_W}}
\newcommand{\trm}{m_{\perp}}
\newcommand{\trp}{p_{\perp}}
\newcommand{\trmm}{m_{\perp}^2}
\newcommand{\trpp}{p_{\perp}^2}
\newcommand{\alp}{\alpha_s}

\newcommand{\alps}{\alpha_s}
\newcommand{\sqrts}{$\sqrt{s}$}
\newcommand{\LO}{$O(\alpha_s^0)$}
\newcommand{\Oa}{$O(\alpha_s)$}
\newcommand{\Oaa}{$O(\alpha_s^2)$}
\newcommand{\PT}{p_{\perp}}
\newcommand{\JPSI}{J/\psi}
\newcommand{\sh}{\hat{s}}
\newcommand{\uh}{\hat{u}}
\newcommand{\MP}{m_{J/\psi}}
\newcommand{\PO}{I\!\!P}
\newcommand{\xbj}{x}
\newcommand{\xpom}{x_{\PO}}
\newcommand{\ttbs}{\char'134}
\newcommand{\xpomlo}{3\times10^{-4}}  
\newcommand{\xpomup}{0.05}  
\newcommand{\dgr}{^\circ}
\newcommand{\WBoson}{\mbox{$W$}}
\newcommand{\fbarn}{\,\mbox{{\rm fb}}}
\newcommand{\fbarnt}{\,\mbox{{\rm fb$^{-1}$}}}
\newcommand{\dsdx}[1]{$d\sigma\!/\!d #1\,$}

\newcommand{\dstar}{\ensuremath{D^*}}
\newcommand{\dstarp}{\ensuremath{D^{*+}}}
\newcommand{\dstarm}{\ensuremath{D^{*-}}}
\newcommand{\dstarpm}{\ensuremath{D^{*\pm}}}
\newcommand{\zDs}{\ensuremath{z(\dstar )}}
\newcommand{\Wgp}{\ensuremath{W_{\gamma p}}}
\newcommand{\ptds}{\ensuremath{p_t(\dstar )}}
\newcommand{\etads}{\ensuremath{\eta(\dstar )}}
\newcommand{\ptj}{\ensuremath{p_t(\mbox{jet})}}
\newcommand{\ptjn}[1]{\ensuremath{p_t(\mbox{jet$_{#1}$})}}
\newcommand{\etaj}{\ensuremath{\eta(\mbox{jet})}}
\newcommand{\detadsj}{\ensuremath{\eta(\dstar )\, \mbox{-}\, \etaj}}
%
%
\newcommand{\pbarnt}{\,\mbox{{\rm pb$^{-1}$}}}
\newcommand{\gev}{\,\mbox{GeV}}
\newcommand{\pb}{\,\rm pb}
\newcommand{\qsq}{\ensuremath{Q^2} }
\newcommand{\gevsq}{\ensuremath{\mathrm{GeV}^2} }
\newcommand{\et}{\ensuremath{E_t^*} }
\newcommand{\rap}{\ensuremath{\eta^*} }
\newcommand{\gp}{\ensuremath{\gamma^*}p }
\newcommand{\dsiget}{\ensuremath{{\rm d}\sigma_{ep}/{\rm d}E_t^*} }
\newcommand{\dsigrap}{\ensuremath{{\rm d}\sigma_{ep}/{\rm d}\eta^*} }
\newcommand{\eV}{\mbox{e\hspace{-0.08em}V}}

\def\Journal#1#2#3#4{{#1} {\bf #2} (#3) #4}
\def\NCA{\em Nuovo Cimento}
\def\NIM{\em Nucl. Instrum. Methods}
\def\NIMA{{\em Nucl. Instrum. Methods} {\bf A}}
\def\NPB{{\em Nucl. Phys.}   {\bf B}}
\def\PLB{{\em Phys. Lett.}   {\bf B}}
\def\PRL{\em Phys. Rev. Lett.}
\def\PRD{{\em Phys. Rev.}    {\bf D}}
\def\ZPC{{\em Z. Phys.}      {\bf C}}
\def\EJC{{\em Eur. Phys. J.} {\bf C}}
\def\CPC{\em Comp. Phys. Commun.}

\begin{titlepage}

\noindent
\begin{flushleft}
{\tt DESY 09-032    \hfill    ISSN 0418-9833} \\
{\tt November 2009} \\
\end{flushleft}

\vspace{2cm}
\begin{center}
\begin{Large}

{\bf Jet Production in {\boldmath $ep$} Collisions at High {\boldmath $Q^2$} \\ and Determination of {\boldmath $\alpha_s$}}

\vspace{2cm}

H1 Collaboration

\end{Large}
\end{center}

\vspace{2cm}

\begin{abstract}

The production of jets is studied in deep-inelastic $e^{\pm}p$ scattering at large negative four momentum transfer squared $150<Q^2<15000$~GeV$^2$ using HERA data taken in 1999-2007, corresponding to an integrated luminosity of $395\pbarnt$. Inclusive jet, 2-jet and 3-jet cross sections, normalised to the neutral current deep-inelastic scattering cross sections, are measured as functions of $Q^2$, jet transverse momentum and proton momentum fraction. The measurements are well described by perturbative QCD calculations at next-to-leading order corrected for hadronisation effects. The strong coupling as determined from these measurements is \mbox{$\alpha_s(M_Z) = 0.1168 ~\pm 0.0007 \,\mathrm{(exp.)}~ ^{+0.0046}_{-0.0030}\,\mathrm{(th.)}~ \pm 0.0016\,$({\scshape pdf})}.

\end{abstract}

\vspace{1.5cm}

\begin{center}
Accepted by \EJC
\end{center}

\end{titlepage}

%
%
%
\begin{flushleft}

F.D.~Aaron$^{5,49}$,           
C.~Alexa$^{5}$,                
K.~Alimujiang$^{11}$,          
V.~Andreev$^{25}$,             
B.~Antunovic$^{11}$,           
A.~Asmone$^{33}$,              
S.~Backovic$^{30}$,            
A.~Baghdasaryan$^{38}$,        
E.~Barrelet$^{29}$,            
W.~Bartel$^{11}$
% ...

\bigskip{\it
% ...
$ ^{2}$ Vinca  Institute of Nuclear Sciences, Belgrade, Serbia \\
% ...
$ ^{5}$ National Institute for Physics and Nuclear Engineering (NIPNE) ,
          Bucharest, Romania \\
% ...
$ ^{11}$ DESY, Hamburg, Germany \\
% ...
$ ^{25}$ Lebedev Physical Institute, Moscow, Russia$^{ e}$ \\
% ...
$ ^{29}$ LPNHE, Universit\'{e}s Paris VI and VII, IN2P3-CNRS,
          Paris, France \\
 $ ^{30}$ Faculty of Science, University of Montenegro,
          Podgorica, Montenegro$^{ e}$ \\
% ...
$ ^{33}$ Dipartimento di Fisica Universit\`a di Roma Tre
          and INFN Roma~3, Roma, Italy \\
% ...
$ ^{38}$ Yerevan Physics Institute, Yerevan, Armenia \\
% ...
$ ^{49}$ Also at Faculty of Physics, University of Bucharest,
          Bucharest, Romania \\
% ...
$ ^e$ Supported by the Deutsche Forschungsgemeinschaft \\
% ...
}
\end{flushleft}
%

\newpage

\section{Introduction}

\begin{figure}[h]
\centering
 \includegraphics[height=4.4cm]{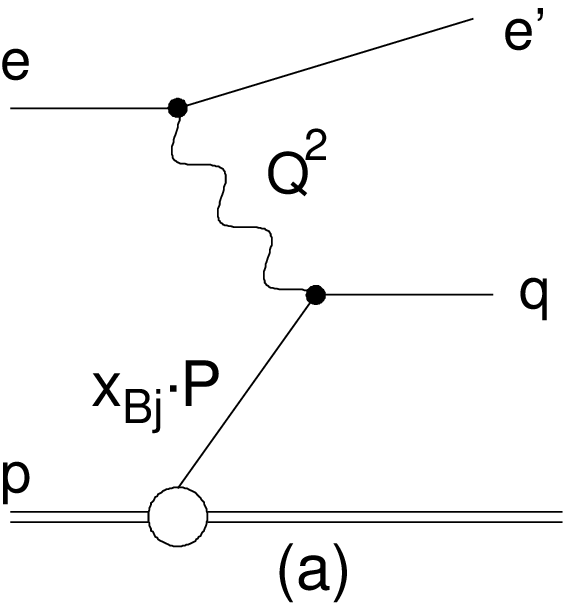}\hskip1.0cm
 \includegraphics[height=4.4cm]{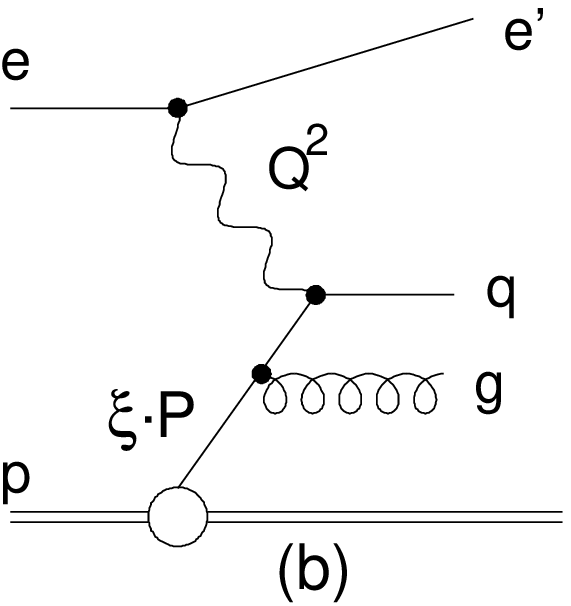}\hskip1.0cm
 \includegraphics[height=4.4cm]{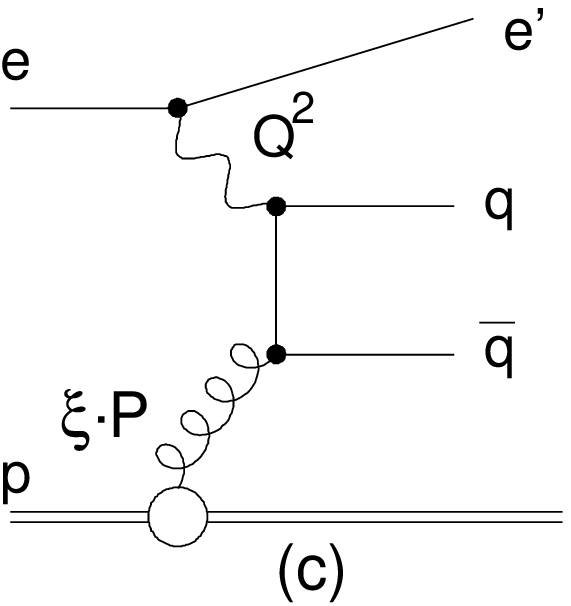}
\caption[Diagrams of different order in $\alpha_s$ in deep-inelastic
lepton-proton scattering]%
{
Deep-inelastic lepton-proton scattering at different orders in
$\alpha_s$: (a) Born contribution $\mathcal{O}(1)$, (b) example of the QCD Compton
    scattering $\mathcal{O}(\alpha_s)$ and (c) boson-gluon fusion $\mathcal{O}(\alpha_s)$.
}
\label{fig:feynborn}
\end{figure}

	Jet production in neutral current (NC) deep-inelastic scattering (DIS) at HERA provides an
important testing ground for Quantum Chromodynamics (QCD). While inclusive DIS gives only indirect information on the strong coupling via scaling violations of the proton structure functions, the production of jets allows a direct measurement of $\alpha_s$. The Born level contribution to DIS (figure \ref{fig:feynborn}a) generates no transverse momentum in the Breit frame, where the virtual boson and the proton collide head on \cite{BreitFrame}. Significant transverse momentum $P_T$ in the Breit frame is produced at leading order (LO) in the strong coupling $\alpha_s$ by the QCD-Compton (figure \ref{fig:feynborn}b) and boson-gluon fusion  (figure \ref{fig:feynborn}c) processes. 

In leading order the proton's momentum fraction carried by the
emerging parton is given by $\xi =  x_{\rm Bj} ( 1+ M^2_{12} / Q^2
)$. The variable $x_{\rm Bj}$ denotes the Bjorken scaling variable,
$M_{12}$ the invariant mass of two jets of highest $P_T$ and $Q^2$ 
the negative four momentum transfer squared. In the kinematical
regions of low $Q^2$, low $P_T$ and low $\xi$, boson-gluon fusion
dominates the jet production and provides direct sensitivity to the
gluon component of proton density functions (PDFs)
\cite{Adloff:2000tq}. 

Analyses of inclusive jet production in DIS at high $Q^2$ were
previously performed by the H1 \cite{H1Incl:2007pb} and ZEUS
\cite{Chekanov:2006yc} collaborations at HERA. These analyses are
based on data taken during 1999 and 2000 (HERA-I) and use jet
observables to test the running of the strong coupling and extract its
value at the $Z^0$ boson mass. In this paper an integrated luminosity
six times larger than available in the previous H1
analysis\cite{H1Incl:2007pb} is used. The ratios of jet cross sections
to the corresponding NC DIS cross sections, henceforth referred to as
normalised jet cross sections, are measured. These ratios benefit from
a partial cancellation of experimental and 
theoretical uncertainties. The measurements are compared with
perturbative QCD (pQCD) predictions at next-to-leading order (NLO)
corrected for hadronisation effects, and $\alpha_s$ is extracted from
a fit of the predictions to the data. The measurements presented in
this paper supersede the previously published normalised jet cross
sections in \cite{H1Incl:2007pb}.

\section{Experimental Method}
\label{sec:expmethod}

	The data sample was collected with the H1 detector at HERA in the years 1999 to 2007 when HERA collided electrons or positrons\footnote{Unless otherwise stated, the term "electron" is used in the following to refer to both electron and positron.}
of energy $E_e = 27.6$~GeV with protons of energy $E_p = 920\gev$, providing a centre-of-mass energy $\sqrt{s}=319\gev$. The data sample used in this analysis corresponds to an integrated luminosity of $395\pbarnt$, comprising $153\pbarnt$ recorded in $e^-p$ collisions and $242\pbarnt$ in $e^+p$ collisions.

\subsection{H1 detector}

	A detailed description of the H1 detector can be found in~\cite{Abt:1996hi,Appuhn:1996na}. H1 uses a right-handed coordinate system with the origin at the nominal interaction point and the $z$-axis along the beam direction. The positive $z$ direction, also called the forward direction, is given by the outgoing proton beam. Polar angles $\theta$ and azimuthal angles $\phi$ are defined with respect to this axis.
The pseudorapidity is related to the polar angle  $\theta$ by $\eta=-\mathrm{ln}\,\mathrm{tan}(\theta/2)$. The detector components important for this analysis are described below.

The electromagnetic and hadronic energies are measured using
the Liquid Argon (LAr) calorimeter in the polar angular range
$4^\circ <\theta <154^\circ$ and with full azimuthal coverage
\cite{h1cal}. The LAr calorimeter consists of an
electromagnetic section ($20$~to~$30$ radiation lengths) with
lead absorbers and a hadronic section with steel 
absorbers. The total depth of the LAr calorimeter varies between 
$4.5$ and $8$ hadronic interaction lengths. The energy 
resolution is $\sigma_E/E = 12\%/\sqrt{E \;/\gev}\oplus 1\%$ 
for electrons and $\sigma_{E}/E = 50\%/\sqrt{E\;/\gev}\oplus 
2\%$ for hadrons, as obtained from test beam
measurements~\cite{Andrieu:1993tz}. In the backward region
($153\dgr\le\theta\le 177\dgr$) energy is measured by a
lead/scintillating fibre Spaghetti-type Calorimeter (SpaCal)
composed of an electromagnetic and a hadronic section
\cite{Appuhn:1996na}. The central tracking system
($20\dgr\le\theta\le 160\dgr$) is located inside the LAr
calorimeter and consists of drift and proportional chambers,
complemented by a silicon vertex detector covering the range
$30\dgr\le\theta\le 150\dgr$~\cite{Pitzl:2000wz}. The chambers
and calorimeters are surrounded by a superconducting solenoid
providing a uniform field of $1.16\,\mathrm{T}$ inside the
tracking volume. The luminosity is determined by measuring the
event rate of the Bethe-Heitler process (\mbox{$ep\rightarrow
ep\gamma$}), where the photon is detected in a calorimeter
close to the beam pipe at $z=-103\,\mathrm{m}$.
	                                                                                                          
\subsection{Event and jet selection}\label{sect:Sel}

The NC DIS events are triggered and selected by requiring a compact
\mbox{energy} deposit in the electromagnetic part of the LAr
calorimeter. The scattered electron is identified as the isolated
cluster of highest transverse momentum \cite{Adloff:2003uh}. Its
reconstructed energy is requested to exceed \mbox{11 GeV}. Only the
regions of the calorimeter 
where the trigger efficiency is greater than $98\%$ are used for the
detection of the scattered electron. These requirements ensure that
the overall trigger efficiency reaches $99.5\%$. In the central
region, $30\dgr\leq \theta \leq 155\dgr$, the cluster has to be
associated with a track measured in the inner tracking chambers and
matched to the primary event vertex. The \mbox{$z$-coordinate} of the
primary event vertex is required to be within $\pm 35~\rm cm$ of the
nominal position of the interaction point.
	
The remaining clusters in the calorimeters and charged tracks are
attributed to the hadronic final state, which is reconstructed using
an energy flow algorithm that avoids double counting of energy
\cite{Peez:2003zd, Portheault:2005uu}. Electromagnetic and hadronic
energy calibration and the alignment of the H1 detector are performed
following the same procedure as in \cite{Adloff:2003uh}. The total
longitudinal energy balance, calculated as the difference of the total
energy $E$ and the longitudinal component of the total momentum $P_z$,
calculated from all detected particles including the scattered
electron, must satisfy $35 < E -P_{z} < 65~\gev$. This requirement
reduces contributions of DIS events with hard initial state photon
radiation. For the latter events, the undetected photons propagating
in the negative $z$ direction lead to values of this observable
significantly lower than the expected value of
$2E_e=55.2\gev$. 
The $E -P_{z}$ requirement together with the scattered electron
selection also reduces contributions from photoproduction, estimated
using Monte Carlo simulations. 
Cosmic muon and beam induced background is reduced 
to a negligible level after combining these cuts with the primary
event vertex selection. 
Elastic QED
Compton and lepton pair production processes are suppressed by
rejecting events containing additional isolated electromagnetic
deposits and low hadronic calorimeter activity.

The kinematical range of this analysis is defined by
\begin{center}
$150 < Q^2 < 15000$~GeV$^2$~~~{\rm and}~~~$0.2 < y < 0.7$\,,
\end{center}
where $y=Q^2/(s\,x_{\rm Bj})$ quantifies the inelasticity of the interaction. 
These two variables are reconstructed from the four momenta of the
scattered electron and the hadronic final state particles using the
electron-sigma method \cite{Bassler:1994uq}. The selection of events
passing all the above cuts is the NC DIS sample, which forms the basis of 
the subsequent analysis.

The jet finding is performed in the Breit frame, where the boost from the laboratory system is determined by $Q^2$, $y$ and by the azimuthal angle $\phi_e$ of the scattered electron. Particles
of the hadronic final state are clustered into jets using the inclusive $k_T$ algorithm \cite{Ellis:1993tq} with the massless
$P_T$ recombination scheme and with the distance parameter $R_0 = 1$ in the $\eta-\phi$ plane. The cut $-0.8 <
\eta_{\rm Lab}^{\rm jet} < 2.0$, where $\eta_{\rm Lab}^{\rm jet}$ is the jet pseudorapidity in the laboratory frame, ensures that jets are contained within the acceptance of the LAr calorimeter and are well calibrated.

	Jets are ordered by decreasing transverse momentum $P_{T}$ in the Breit 
frame, which is identical to the transverse energy $E_T$ for massless jets. 
The jet with highest $P_T$ is referred to as the "leading jet". Every jet with 
the transverse momentum $P_{T}$ in the Breit frame satisfying 
$7 < P_{T} < 50\gev$ contributes to the inclusive jet cross section. 
The upper cutoff is necessary for the integration of the NLO calculation.
The steeply falling transverse momentum spectrum leaves almost no jets 
above 50 GeV.
Events with at least two (three) jets with transverse momentum 
$5 < P_{T} < 50\gev$ are considered as 2-jet (3-jet) events. In order to avoid 
regions of phase-space where fixed order perturbation theory is not 
reliable \cite{Frixione:1997ks}, 2-jet events are accepted only if the 
invariant mass $M_{12}$ of the two leading jets exceeds $16\gev$. The same 
requirement, $M_{12}>16\gev$, is applied to the 3-jet events so that the 3-jet 
sample is a subset of the 2-jet sample.
	
	After this selection, the inclusive jet sample contains a total of 143811 jets in 104014 events. The 2-jet sample contains 47278 events and the 3-jet sample 7054 events. 

\subsection{Definition of the observables}

The measurements presented in this paper refer to the phase-space
given in table \ref{tab::PhaseSpace}. Normalised inclusive jet cross
sections are measured as functions of $Q^2$ and double differentially
as function of $Q^2$ and the transverse jet momentum $P_T$ in the
Breit frame. Normalised 2-jet and 3-jet cross sections are presented
as a function of $Q^2$. In addition the 2-jet cross sections are
measured double differentially as function of $Q^2$ and the average
transverse momentum of the two leading jets $\left\langle P_T
\right\rangle = \frac{1}{2}\cdot ( P_{T}^{\rm jet1}+P_{T}^{\rm jet2}
)$ or as function of $Q^2$ and of the proton momentum fraction 
$\xi$. The 3-jet cross section normalised to the 2-jet cross section as
function of $Q^2$ is also presented.

The normalised jet cross sections are defined as the ratio of the
differential inclusive jet, 2-jet and 3-jet cross sections to
the differential NC DIS cross section in a given $Q^2$ bin,
multiplied by the respective bin width $W$ in case of a double
differential measurement as indicated by the following equations:
\begin{eqnarray}\label{eqn}
\frac{\sigma_{\rm jet}}{\sigma_{\rm NC}}\left(Q^2,\, P_T\right) \hspace{1pc} &=& 
\frac{ \text{d}^2\sigma_{\rm jet} / \text{d}Q^2\,\text{d}P_{T} }
     { \text{d}\sigma_{\rm NC}   /  \text{d}Q^2}
     \cdot W(P_T)\\
\frac{\sigma_{\textnormal{2-jet}}}{\sigma_{\rm NC}}\left(Q^2,\, \left\langle P_T\right\rangle\right) &=& 
\frac{ \text{d}^2\sigma_{\textnormal{2-jet}} / \text{d}Q^2\,\text{d}\left\langle P_T\right\rangle }
     { \text{d}\sigma_{\rm NC}   /  \text{d}Q^2}\cdot W(\left\langle P_T\right\rangle)\\
\frac{\sigma_{\textnormal{2-jet}}}{\sigma_{\rm NC}}\left(Q^2,\, \xi \right)  \hspace{1.5pc} &=& 
\frac{ \text{d}^2\sigma_{\textnormal{2-jet}} / \text{d}Q^2\,\text{d}\xi }
     { \text{d}\sigma_{\rm NC}   /  \text{d}Q^2}\cdot W(\xi)
%
\end{eqnarray}	

	The normalised inclusive jet cross section can be viewed as the average jet multiplicity in a given $Q^2$ region and the normalised multi-jet cross sections as multi-jet event rates.

\subsection{Determination of normalised cross sections}
\label{sec:datacorrrection}

In each analysis bin the normalised jet cross section is determined as 
\begin{eqnarray}
\frac{\sigma_{\rm J}}{\sigma_{\rm NC}} = \frac{N_{\rm J}}{N_{\rm NC}} \cdot C\,.
\end{eqnarray}	
Here $N_{\rm J}$ denotes the number of inclusive jets or the number of
2-jet or 3-jet events, respectively, while $N_{\rm NC}$ represents the number 
of NC DIS events in that bin. The bin dependent correction factor $C$ 
takes into account the limited detector acceptance and resolution. The 
correction factors are determined from Monte Carlo simulations as the 
ratio of the normalised jet cross sections obtained from particles at 
the hadron level to the normalised jet cross sections calculated using 
reconstructed particles. 

\begin{figure}[p]
\begin{center}
\epsfig{file=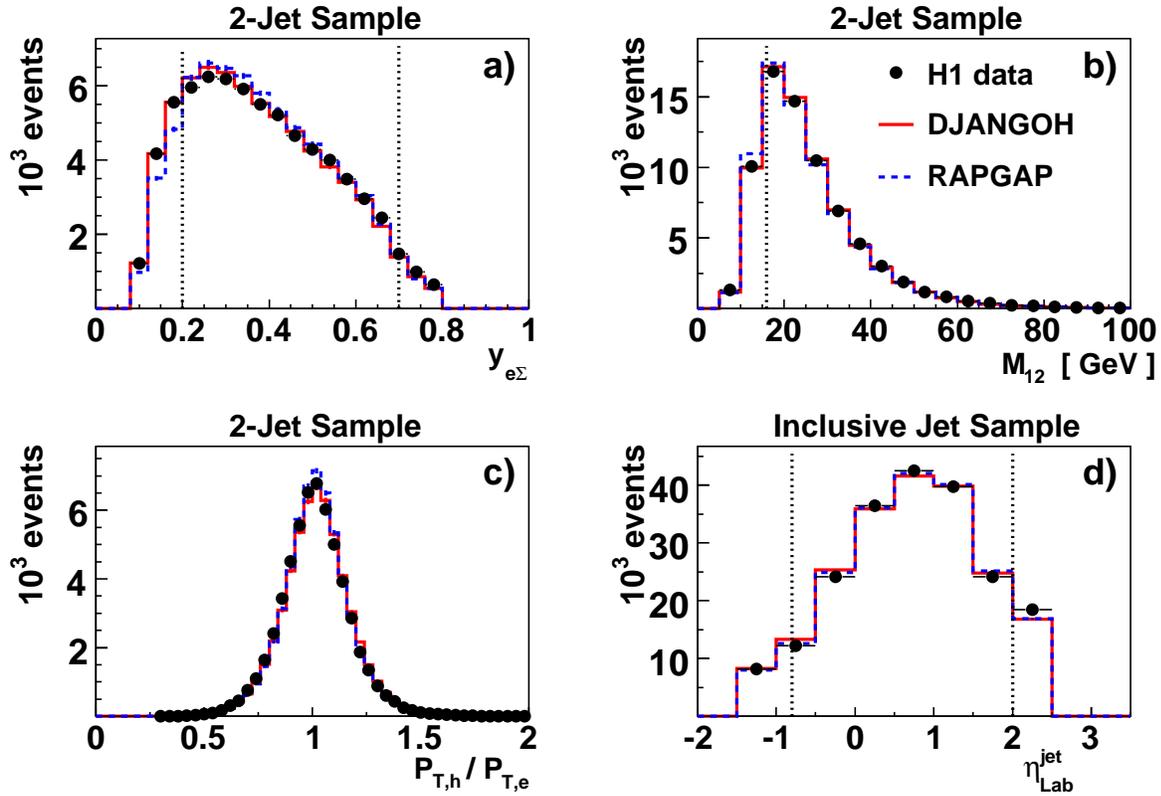,
width=160mm,angle=0,clip= }
\end{center}
\caption{\label{fig::ControlPlots} Distribution of the selected events
  (solid dots) shown as a function of selection variables in an
  extended domain: the inelasticity $y$
  reconstructed with the electron-$\Sigma$ method of 2-jet events (a);
  the invariant mass of the two leading jets $M_{12}$ (b); the
  transverse momentum ratio in the laboratory frame $P_{T,
    h}/P_{T, e}$ of 2-jet events (c); the $\eta_{\rm Lab}^{\rm
    jet}$ of the inclusive jets (d). The data are compared with
  weighted MC simulations, DJANGOH (solid line) and RAPGAP (dashed
  line). Vertical dashed lines indicate the positions of kinematical cuts.}
\end{figure}

The following LO Monte Carlo event generators are used for the
correction procedure: DJANGOH\cite{Charchula:1994kf}, which uses the
Color Dipole Model with QCD matrix element corrections as implemented
in ARIADNE\cite{Lonnblad:1992tz}, and RAPGAP\cite{Jung:1993gf}, based
on QCD matrix elements matched with parton showers in leading log
approximation. In both Monte Carlo generators the hadronisation is
modelled with Lund string fragmentation \cite{Andersson:1983ia}. All
generated events are passed through a GEANT3~\cite{Brun:1987ma} based
simulation of the H1 apparatus and are reconstructed using the same
program chain as for the data. Both RAPGAP and DJANGOH provide a good 
overall description of the inclusive DIS sample. To further improve the
agreement between Monte Carlo and data for the jet samples, the Monte
Carlo events are weighted as a function of $Q^2$ and $y$ and as
function of $P_T$ and $\eta$ of the leading jet in the Breit
frame. In addition, they are weighted as a function of $P_T$ of the
second and third jets when present \cite{Gouzevitch:2008zz}. After
weighting, the simulations provide a good description of the shapes of
all data distributions, some of which are shown in figure
\ref{fig::ControlPlots}.

The binnings in $Q^2$, $P_T$ and $\xi$ used to measure the jet
observables are given in table \ref{tBinning}. The associated bin
purities, defined as the fraction of the events reconstructed in a
particular bin that originate from that bin on the generator level,
are typically $70\%$ and always greater than $60\%$. The correction
factors deviate typically by less than $20\%$ from unity, but reach
$40\%$ difference from unity in the bin 
\mbox{$5<\left\langle P_T\right\rangle<7~$GeV} for
the 2-jet cross section. Arithmetic means of the correction factors
determined from the reweighted RAPGAP and DJANGOH event samples are
used and half of the difference is assigned as a model uncertainty.
	
The above correction factors include QED radiation and electroweak
effects. The effects of QED radiation, which are typically $5\%$, are
corrected for by means of the HERACLES~\cite{Kwiatkowski:1990es}
program. The LEPTO event generator \cite{Ingelman:1996mq} is used to
correct the $e^{+}p$ and $e^-p$ data for their different electroweak
effects which largely cancel in normalised jet cross sections leaving
them below $3\%$. The resulting pure photon exchange cross sections
obtained from $e^{+}p$ and $e^-p$ data samples are then averaged.	

\subsection{Experimental uncertainties}\label{sUncert}

The systematic uncertainties of the jet observables are determined by propagating the corresponding estimated measurement errors through the full analysis:

\begin{itemize} 
\item The relative uncertainty of the electron energy calibration is
  typically between $0.7\%$ and $1\%$ for most of the events and
  increases up to $2\%$ for electrons in the forward direction. The
  absolute uncertainty of the electron polar angle is \mbox{$3$
    mrad}. Uncertainties in the electron reconstruction affect the
  event kinematics and thus the boost to the Breit frame. This in turn
  leads to a relative error of $0.5\%$ to $1.5\%$ on the normalised
  cross sections for each of the two sources, electron polar angle and energy.

\item The relative uncertainty on the energy of the total
  reconstructed hadronic final state as well as of jets is estimated
  to be $1.5\%$  \cite{Gouzevitch:2008zz}. It is dominated by the
  uncertainty of the hadronic energy scale of the calorimeter. This
  error is estimated using a procedure similar to that used in
  \cite{Adloff:2003uh} based on the transverse momentum conservation
  in the laboratory frame between the hadronic final state $P_{T,h}$
  and the electron $P_{T,e}$. This systematic uncertainty is reduced
  with respect to the previous measurement \cite{H1Incl:2007pb} due to
  the restricted pseudorapidity range in which jets are reconstructed and
  due to the improved statistics in the calibration procedure. The
  hadronic energy scale uncertainty affects mainly the jet cross
  section through the calibration of $P_{T}$ and, to a lesser extent,
  the NC DIS cross section through the reconstruction of $y$. The
  resulting errors range between $1\%$ and $5\%$ and increase up to
  $7\%$ when $P_T$ exceeds 30 GeV. The relative uncertainty due to the
  hadronic energy scale is reduced on average by about $20\%$ for the
  normalised jet cross sections compared to the jet cross sections.

\item The model dependence of the detector correction factors is
  estimated as described in section \ref{sec:datacorrrection}. It
  reflects the sensitivity of the detector simulation to the details
  of the model, especially the parton showering, and their impact on
  the migration between adjacent bins in $P_T$. The model dependence
  ranges typically from $1\%$ to $2\%$ for $P_T$ below $30\gev$ and to
  $4\%$ above, independently of $Q^2$.

\item The uncertainties of the luminosity measurements, the trigger efficiency and the electron identification efficiency cancel in the normalised cross section. In addition, the model dependence of the QED radiative corrections, which is estimated to be $1\%$ \cite{Adloff:2003uh}, is expected to cancel in the normalised cross sections.
\end{itemize}

The statistical errors for the normalised inclusive jet cross section
take into account the statistical correlations which arise because
there can be more than one jet per event \cite{Gouzevitch:2008zz}. The
statistical errors are considerably smaller compared to the previous
HERA-I publication \cite{H1Incl:2007pb}. They are typically between
$1\%$ and $2\%$ for the normalised inclusive and 2-jet cross sections and 
do not exceed 10\% in the regions of high transverse momentum $P_T$ or
high boson virtuality $Q^2$.

The dominant experimental errors on the jet cross sections arise from
the uncertainty on the hadronic energy scale. The second most
important source of systematic errors is the model dependence of the
data correction, which becomes comparable to or exceeds the former in
regions of highest jet $P_T$. The overall experimental error,
calculated as the quadratic sum of all the contributions inventoried
above, ranges typically between $3\%$ and $6\%$, but increases up to
$15\%$ in the regions of highest $P_T$ or $Q^2$, dominated there by
statistical uncertainties. The experimental
errors for normalised cross sections are reduced by $30\%$ up to
$50\%$ compared to those for unnormalised cross sections.

\section{NLO QCD prediction of jet cross sections}\label{sec:nlo}

Reliable quantitative predictions of jet cross sections in DIS require
the perturbative calculations to be performed at least to
next-to-leading order in the strong coupling. By using the inclusive
$k_T$ jet algorithm with radius parameter $R=1$, the observables used
in the present analysis are infrared and collinear safe and the
non-perturbative effects are expected to be small \cite{Adloff:2000tq}.
In addition, applying this algorithm in the Breit
frame has the advantage that initial state singularities can be
absorbed in the definition of the proton parton densities 
\cite{BIB::FACTOR_JET}.

Jet cross sections are predicted at the parton level using the same
jet definition as in the data analysis. The QCD predictions for the
jet cross sections are calculated using the NLOJET++ program at NLO in
the strong coupling \cite{Nagy:2001xb}. The NC DIS cross section is
calculated at $\mathcal{O}(\alpha_s)$ with the DISENT package
\cite{Catani:1996vz}. The FastNLO program~\cite{Kluge:2006xs} provides
an efficient method to calculate these cross sections based on matrix
elements from NLOJET++ and DISENT, convoluted with the PDFs of the 
proton and as a function of $\alpha_s$. The program includes a coherent
treatment of the renormalisation and factorisation scale dependences
of all ingredients to the cross section calculation, namely the matrix
elements, the PDFs and $\alpha_s$.
			
When comparing data and theory predictions the strong
coupling at the $Z^0$ boson mass is taken to be $\alpha_s(M_Z) =
0.1168$ and is evolved as a function of the renormalisation scale with
two loop precision. The calculations are performed in the
$\overline{MS}$ scheme for five massless quark flavours. The PDFs of
the proton are taken from the CTEQ6.5M set \cite{Tung:2006tb}. The
factorisation scale $\mu_f$ is taken to be $Q$ and the renormalisation
scale $\mu_r$ to be $\sqrt{(Q^2 + P_{T\raisebox{0pt}[0pt][0pt]{,\,}\text{obs}}^{2})/2}$ for the NLO
predictions, with $P_{T,\text{obs}}$ denoting $P_T$ for the inclusive
jet, $\left\langle P_T \right\rangle$ for 2-jet and
\mbox{$\frac{1}{3}\cdot ( P_{T}^{\rm jet1}+P_{T}^{\rm jet2}+P_{T}^{\rm
jet3} )$} for the 3-jet cross sections. This choice of the
renormalisation scale is motivated by the presence of two hard scales,
$P_T$ and $Q$ in the jet production in DIS. 
For the calculation of inclusive DIS cross sections, the
renormalisation scale $\mu_r=Q$ is used. 
No QED radiation or $Z^0$
exchange is included in the calculations, but the running of the
electromagnetic coupling with $Q^2$ is taken into account.

Hadronisation corrections are calculated for each bin using Monte
Carlo event generators. These corrections are determined as the ratio
of the cross section at the hadron level to the cross section at the
parton level after parton showers. They typically differ by less than
$10\%$ from unity and are obtained using the event generators DJANGOH
and RAPGAP which agree to within $2\%$ to $4\%$. The arithmetic means of
the two Monte Carlo hadronisation correction factors are used, while
the full difference is considered as systematic error.
	
DJANGOH and RAPGAP both use the Lund string model of 
hadronisation. The analytic calculations carried out in 
\cite{Dasgupta:2007wa} provide an alternative method to estimate the 
effects of hadronisation and to cross-check the hadronisation 
correction procedure described above. They are based on soft gluon 
power corrections and result in a shift of the perturbatively 
calculated spectrum of the inclusive jets:
\begin{eqnarray}
\frac{\textnormal{d}\sigma_{\rm jet}}{\textnormal{d}Q^2\textnormal{d}P_T} \left(P_T \right)\approx \frac{\textnormal{d}\sigma_{\rm jet}^{\rm NLO}}{\textnormal{d}Q^2\textnormal{d}P_T}\left(P_T-\delta \left\langle P_T\right\rangle_{\rm NP} \right)
\end{eqnarray}
The size of the non-perturbative shift $\delta \left\langle
P_T\right\rangle_{\rm NP}$ can be calculated up to one single
non-pertur\-bative parameter
$\alpha_0(\mu_I)=\mu_I^{-1}\int_0^{\mu_I}\alpha_{\rm eff}(k)dk$, which
is the first moment of the effective non-perturbative coupling
$\alpha_{\rm eff}(\mu)$ matched to the strong coupling $\alpha_s(\mu)$
at the scale $\mu_I$. The value of $\alpha_0(\mu_I)$, expected
to be universal \cite{Dokshitzer:1995qm}, was measured to be
$\alpha_0(\mu_I=2\gev)\approx 0.5$ using event shapes observables in DIS by
the H1 Collaboration \cite{Aktas:2005tz}. The hadronisation correction
factors so calculated for the inclusive jet cross section differ in
most of the bins by less than $2\%$ from the average correction factor
obtained from 
DJANGOH and RAPGAP and the maximum difference in all bins does not 
exceed $5\%$ which is within the estimated uncertainty of the hadronisation
correction.

The dominant theoretical error is due to the uncertainty related to
the neglected higher orders in the perturbative calculation. The
accuracy of the NLO calculation is conventionally estimated by
separately varying the chosen scales for $\mu_f$ and $\mu_r$ by
factors in the arbitrary range 0.5 to 2. At high transverse momentum,
above $30$~GeV, the pQCD calculations do not depend monotonically on
$\mu_r$ in some $Q^2$ bins. This happens in the two highest $Q^2$ bins
for the inclusive jet cross section and in six $Q^2$ bins for the
2-jet cross section, where the largest deviation from the central
value is found for factors well inside the range 0.5 to 2. In such
cases the difference between maximum and minimum cross sections found
in the variation interval is taken, in order not to underestimate the
scale dependence. Renormalisation and factorisation scale
uncertainties are added in quadrature, the former outweighing the
latter by a factor of two on average. The uncertainties originating
from the PDFs are estimated using the CTEQ6.5M set of parton 
densities.

	Normalised jet cross sections are calculated by dividing the predicted jet cross sections by the NC DIS cross sections. The renormalisation scale uncertainties are assumed to be uncorrelated between NC DIS and jet cross sections, as well as between 3-jet and 2-jet cross sections for their ratio, whereas the factorisation scale and the parameterisation uncertainty of the PDFs are assumed to be fully correlated.

\section{Results}

In the following, the normalised differential cross sections are presented for inclusive jet, 2-jet and 3-jet production at the hadron level. Tables \ref{tab::jet_Q2} to \ref{tab::2jet_Q2Ksi} and figures \ref{fig::jet_Q2} to \ref{fig::2jet_Q2Ksi} present the measured observables together with their experimental uncertainties and hadronisation correction factors applied to the NLO predictions. These measurements are subsequently used to extract the strong coupling $\alpha_s$ as shown in the table \ref{tab::Fits} and figures \ref{fig::FitIjet1D_MuR} to \ref{fig::FitAlljet1D}.

\subsection{Cross section measurements compared to NLO predictions}

\begin{figure}[p]
\begin{minipage}{19pc}
\begin{flushleft}
\epsfig{file=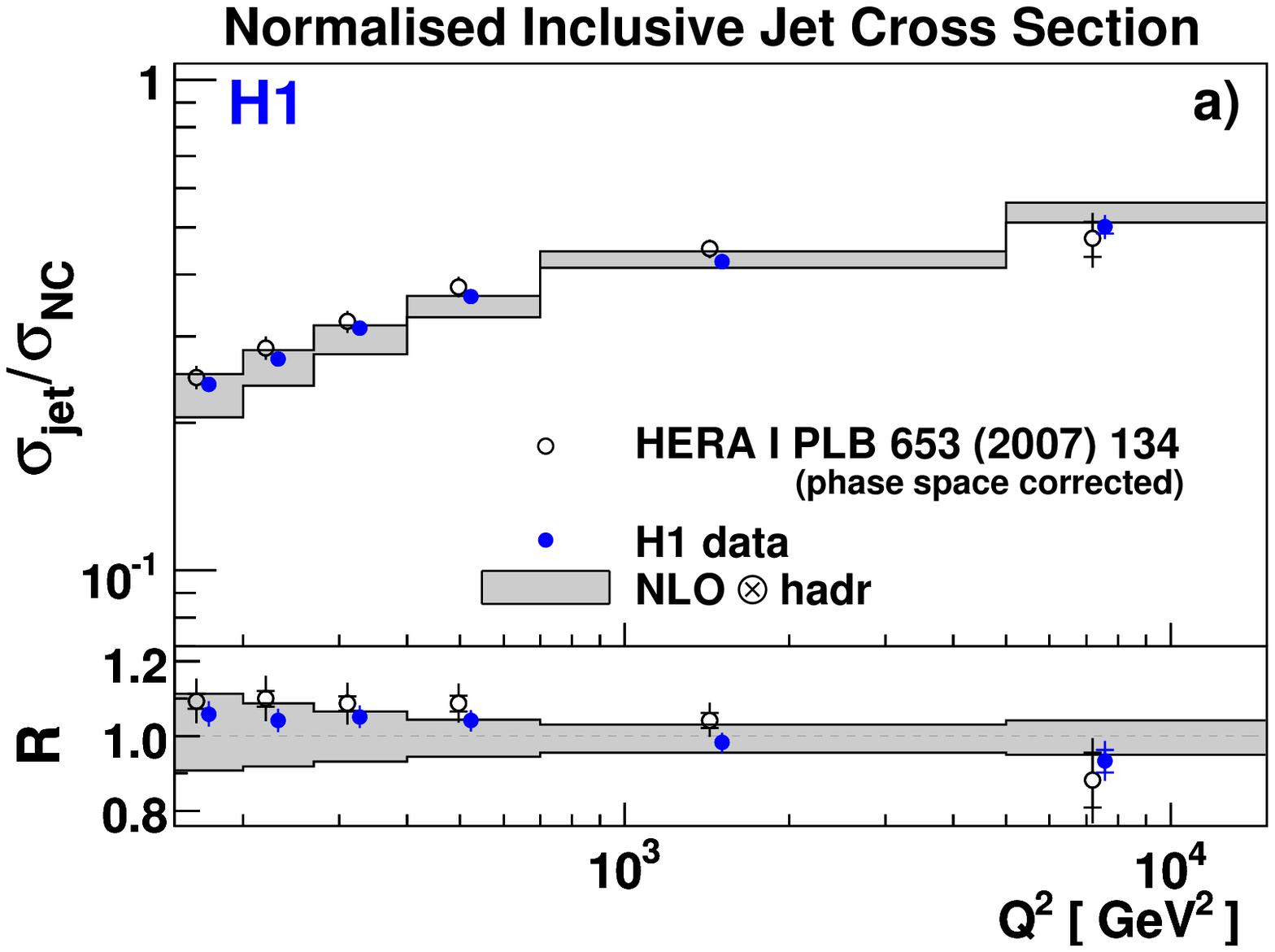,
width=93mm, angle=0, clip= }
\epsfig{file=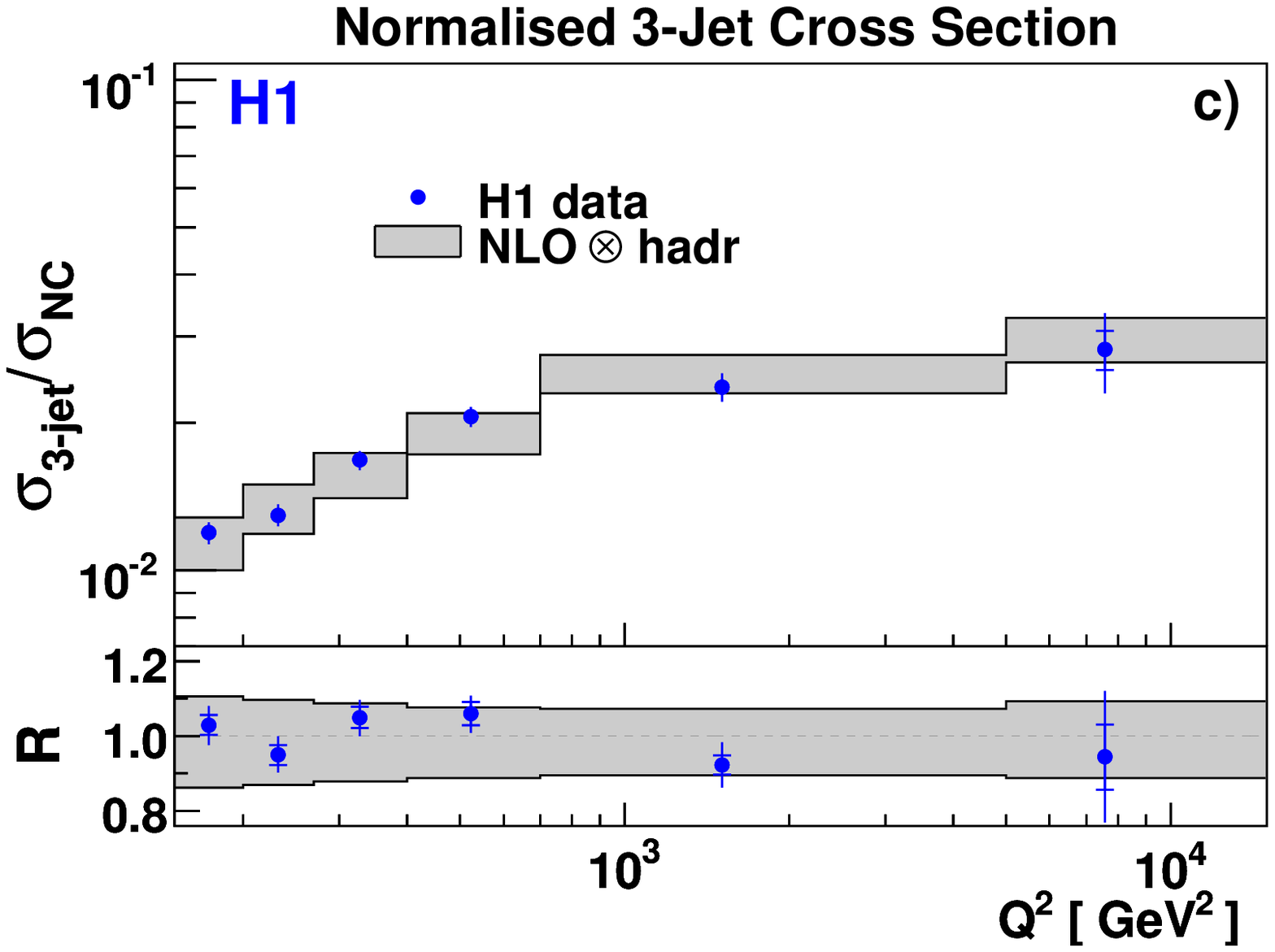,
width=93mm,angle=0,clip= }
\end{flushleft}
\end{minipage}
\begin{minipage}{19pc}
\centering
\epsfig{file=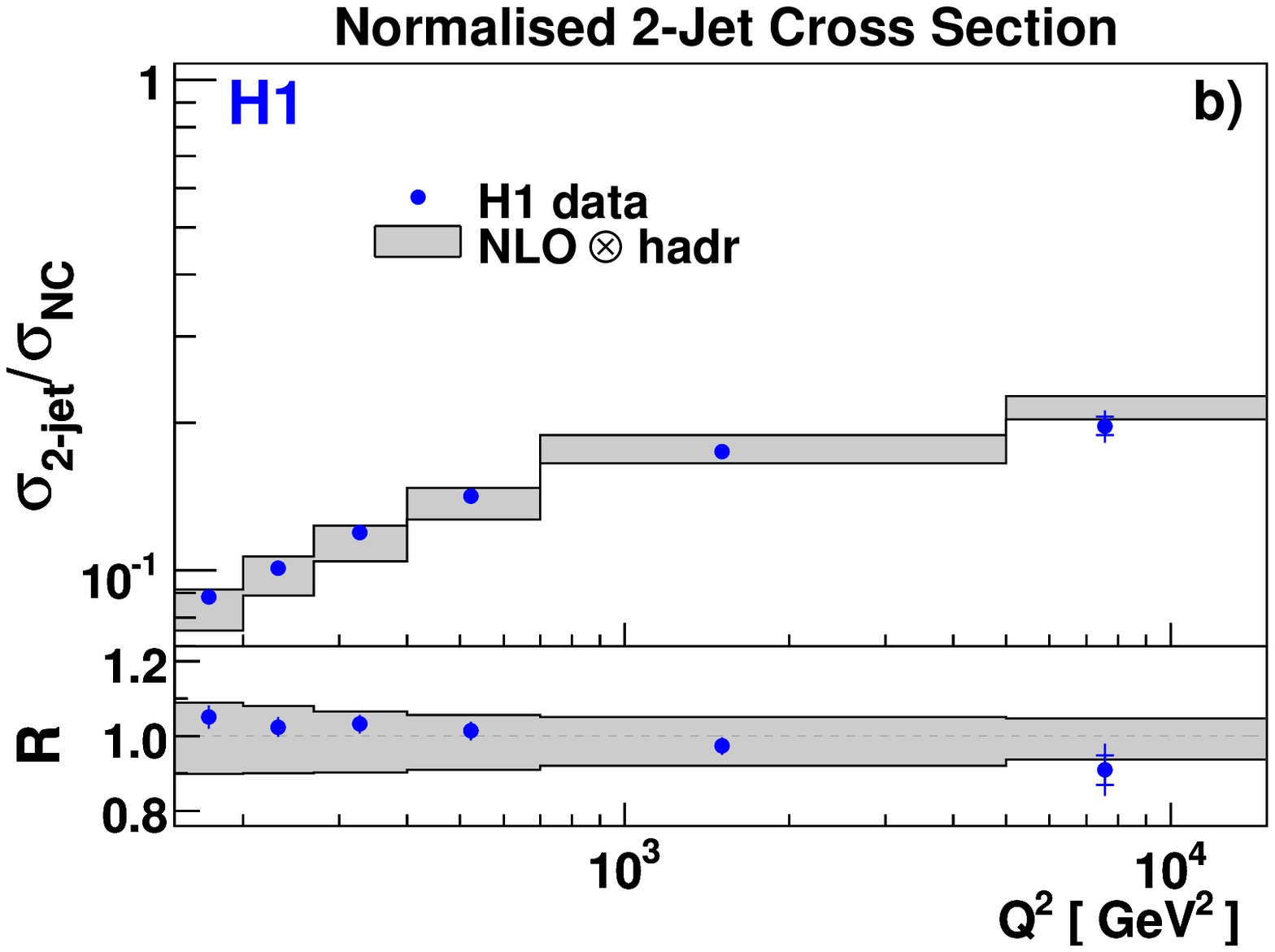,
width=93mm, angle=0,clip= }
\epsfig{file=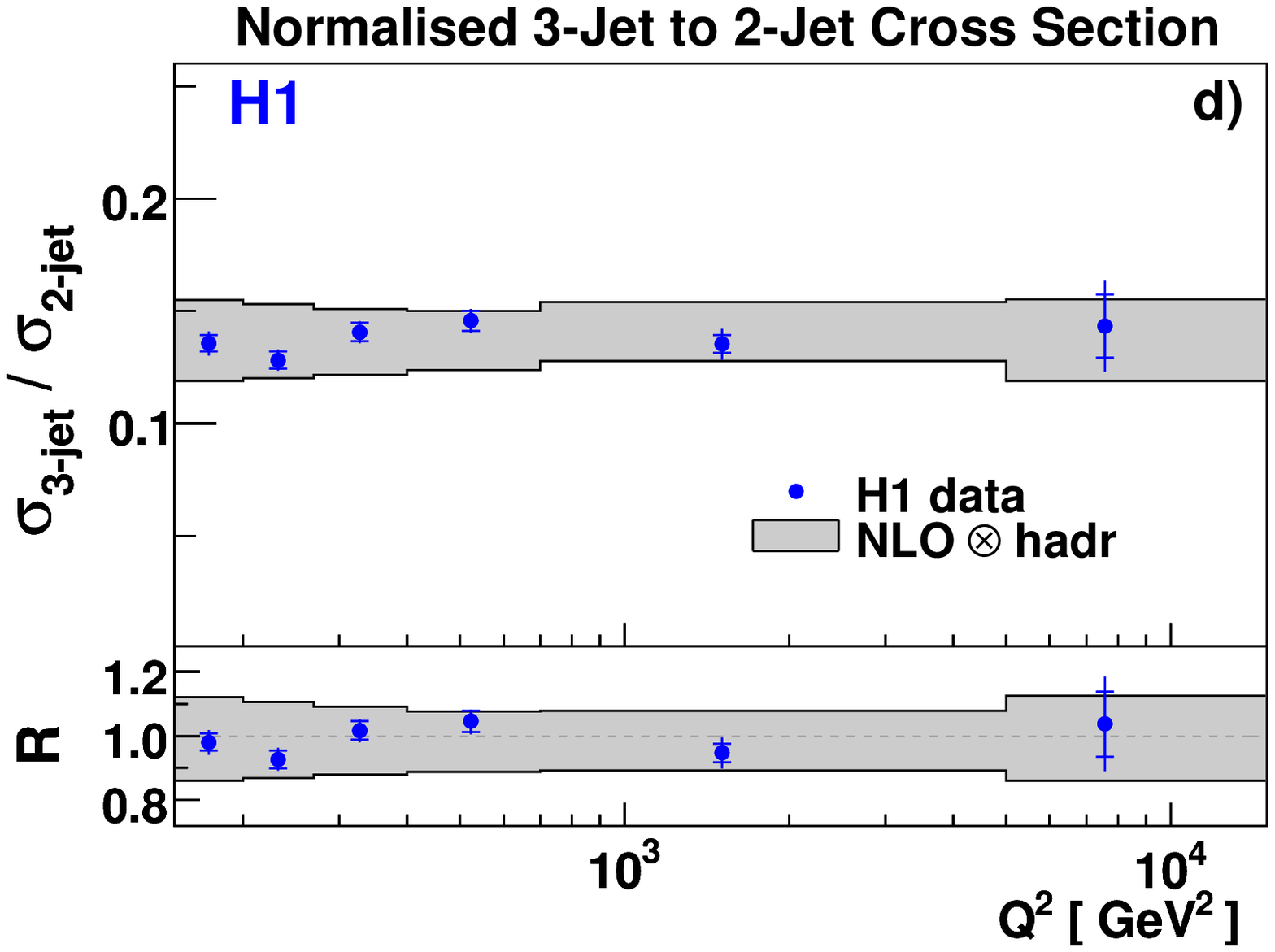,
width=93mm, angle=0,clip= }
\end{minipage}

\caption{\label{fig::jet_Q2}  The normalised inclusive jet (a), 2-jet
  (b) and 3-jet (c) cross sections in NC DIS measured as a function of
  $Q^2$. The ratio of 3-jet to 2-jet cross sections is shown in 
  (d). The measurements refer to the phase-space given in table
  \ref{tab::PhaseSpace}. The points are shown at the average value of
  $Q^2$ within each bin. For the inclusive jets the present data 
  (solid dots) are compared to HERA-I published 
  data~\cite{H1Incl:2007pb}, here shown corrected to the same phase 
  space as the present measurement and shifted in $Q^2$ for clarity 
  (open dots). The inner error bars represent the statistic 
  uncertainties. The outer error bars show the total experimental 
  uncertainties, defined as the quadratic sum of the statistical and 
  systematic uncertainties. The NLO QCD predictions, with parameters 
  described in the section \ref{sec:nlo} and corrected for 
  hadronisation effects are shown together with the theory 
  uncertainties associated with the renormalisation and factorisation
  scales, the PDF and the hadronisation (grey band). The ratio $R$ of
  data with respect to the NLO QCD prediction is shown in the lower 
  part of each plot.}

\end{figure}

\begin{figure}[p]
\begin{center}
\epsfig{file=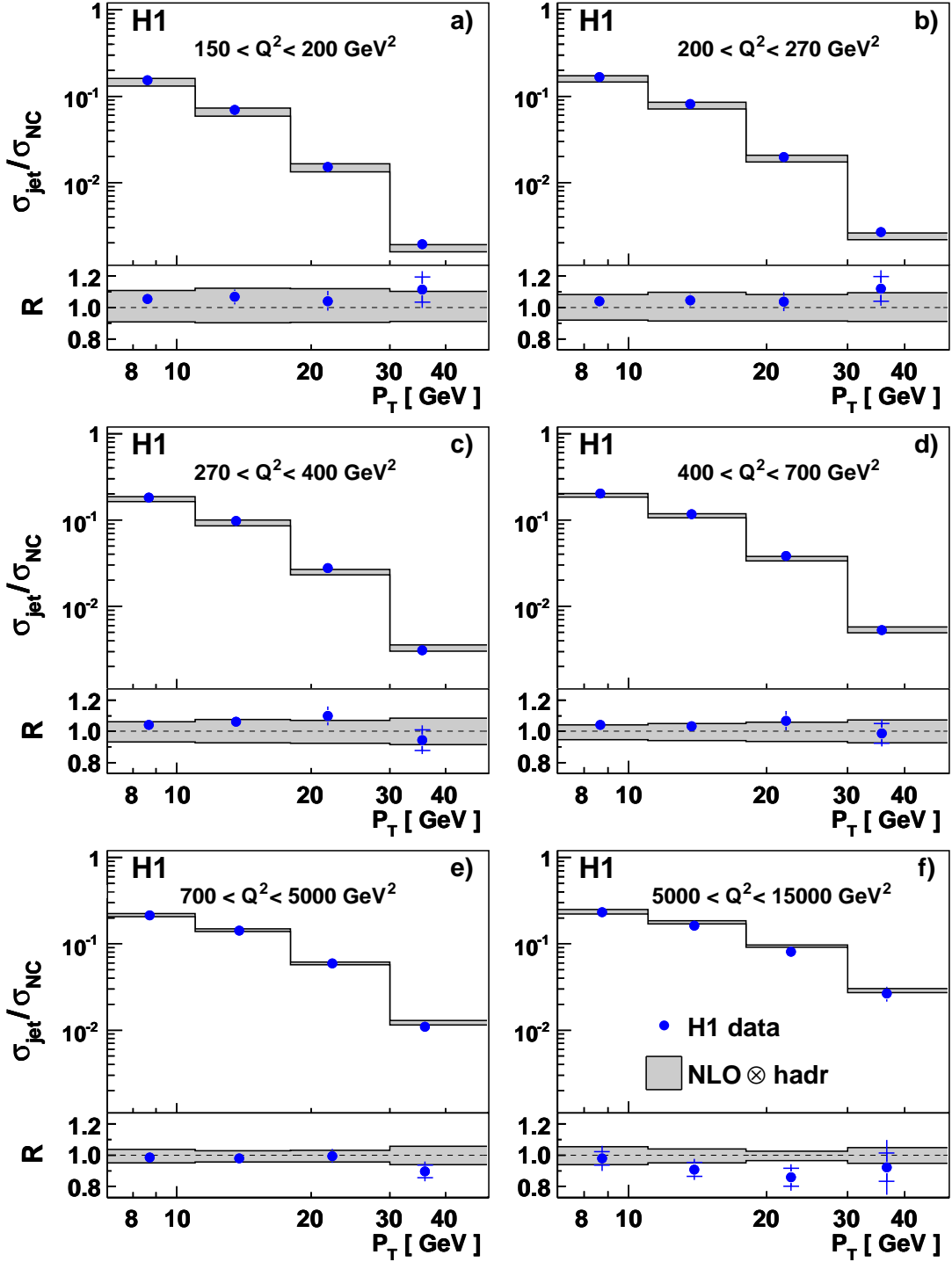,
width=148mm,angle=0,clip= }
\end{center}
\caption{\label{fig::Ijet_Q2ET}The normalised inclusive jet cross
  sections measured as a function of the jet transverse momentum in
  the
  Breit frame $P_{T}$ in regions of $Q^2$. The points are shown at the
  average value of $P_T$ within each bin. Other details are given in 
  the caption to figure \ref{fig::jet_Q2}.}
\end{figure}

\begin{figure}[p]
\begin{center}
\epsfig{file=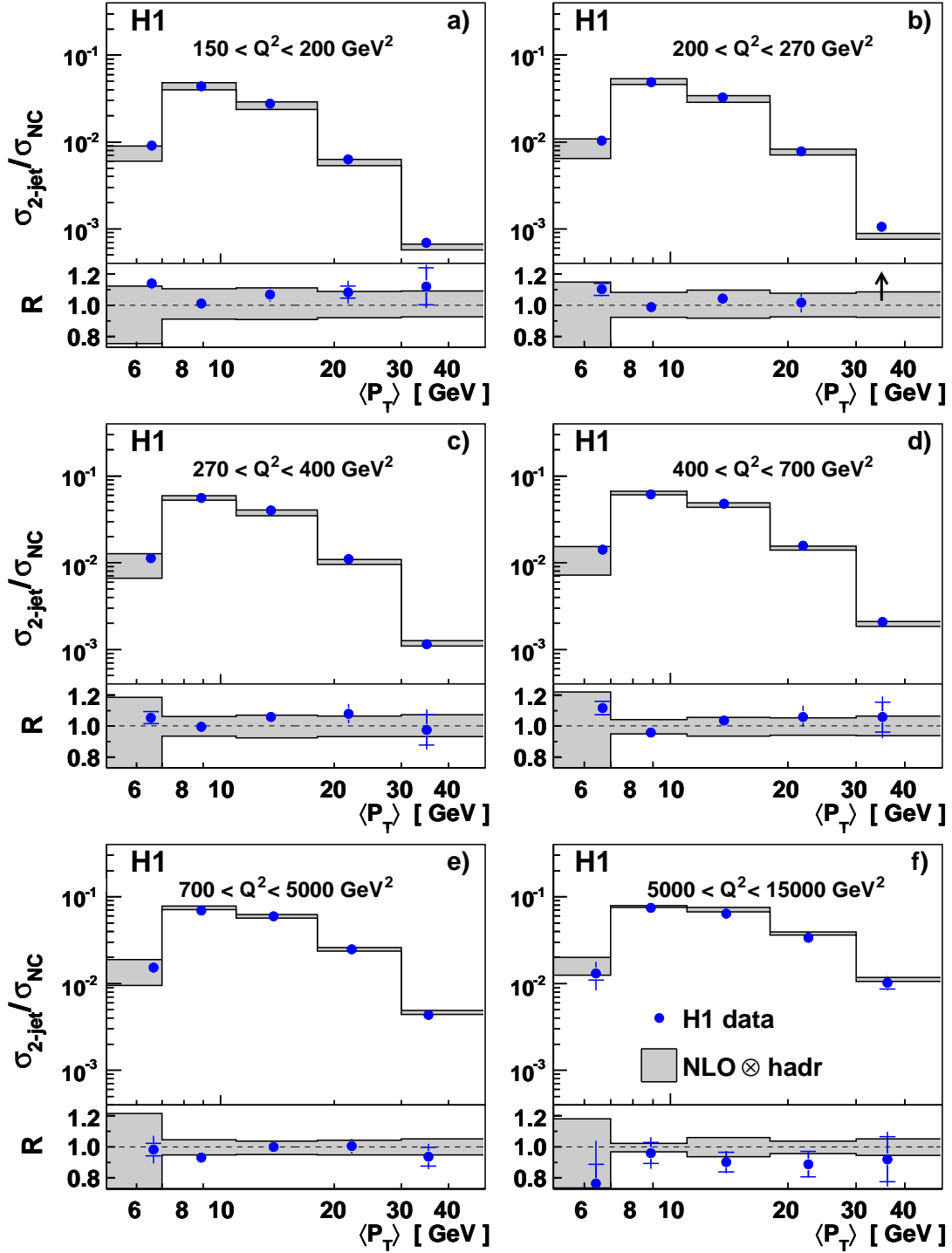,
width=148mm,angle=0,clip= }
\end{center}
\caption{\label{fig::2jet_Q2ET} The normalised 2-jet cross sections
  measured as a function of the average transverse momentum of the two
  leading jets in the Breit frame $\left\langle P_{T}\right\rangle$ in
  regions of $Q^2$. The points are shown at the average value of
  $\left\langle P_{T}\right\rangle$ within each bin. Other details are
  given in the caption to figure \ref{fig::jet_Q2}.}
\end{figure}

\begin{figure}[p]
\begin{center}
\epsfig{file=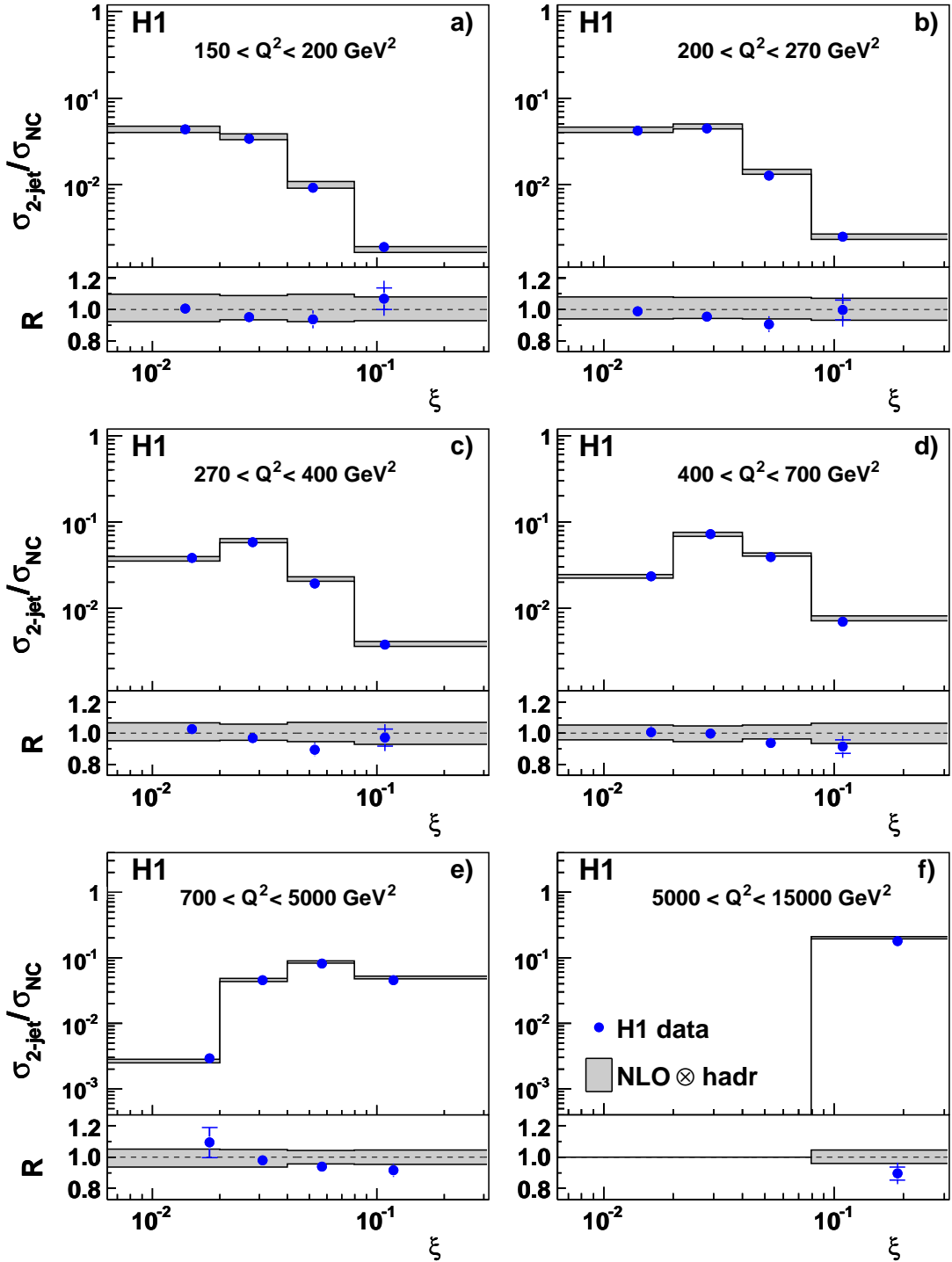,
width=148mm,angle=0,clip= }
\end{center}
\caption{\label{fig::2jet_Q2Ksi} The normalised 2-jet cross sections measured as a function of the proton momentum fraction $\xi$ in regions of $Q^2$. The points are shown at the average value of $\xi$ within each bin. Other details are given in the caption to figure \ref{fig::jet_Q2}.}
\end{figure}

The normalised inclusive jet cross sections as a function of $Q^2$ are shown in figure \ref{fig::jet_Q2}a and \mbox{table \ref{tab::jet_Q2}} together with the NLO predictions and previous measurements by H1 based on HERA-I data \cite{H1Incl:2007pb}. For comparison, the HERA-I data points were corrected for the phase space difference due to the slightly smaller jet pseudorapidity range of the present analysis. The double differential results as a function of $P_T$ in six ranges of $Q^2$ are given in figure \ref{fig::Ijet_Q2ET} and table \ref{tab::Ijet_Q2ET}. Normalised 2-jet (3-jet) cross sections as a function of $Q^2$ and their comparison to NLO are also shown in figure \ref{fig::jet_Q2}b (\ref{fig::jet_Q2}c) and table \ref{tab::jet_Q2}, while the ratio 3-jet to 2-jet is shown in figure \ref{fig::jet_Q2}d. Figures \ref{fig::2jet_Q2ET}, \ref{fig::2jet_Q2Ksi} and tables \ref{tab::2jet_Q2ET}, \ref{tab::2jet_Q2Ksi} present the normalised 2-jet cross section as a function of $\left\langle P_T \right\rangle$ and $\xi$ in six ranges of $Q^2$. 

The new measurement of the normalised inclusive jet cross section is
compatible with the previous H1 data. The precision is improved by
typically a factor of two, as can be seen for example in figure
\ref{fig::jet_Q2}a. The QCD NLO predictions for all normalised jet
cross sections provide a good description of the data over the whole
phase space. In almost all bins the theory error, dominated by the
$\mu_r$ scale uncertainty, is significantly larger than the total
experimental uncertainty, which is dominated by the hadronic energy
scale uncertainty.

The normalised inclusive jet cross section, which may be interpreted
as the average jet multiplicity produced in NC DIS, increases with
$Q^2$ as the available phase space opens (figure \ref{fig::jet_Q2}a)
as do the 2-jet and \mbox{3-jet} rates (figure \ref{fig::jet_Q2}b and
\ref{fig::jet_Q2}c). As $Q^2$ increases, the $P_T$ jet spectra become
harder as can be seen in figure \ref{fig::Ijet_Q2ET} and
\ref{fig::2jet_Q2ET}. The 3-jet rate is observed to be nearly seven
times smaller than the 2-jet rate as shown in figure
\ref{fig::jet_Q2}d. The 2-jet rates measured as a function of $Q^2$
and the momentum fraction $\xi$ are well described by the NLO
calculations (figure \ref{fig::2jet_Q2Ksi}). Kinematic constraints 
from the considered $y$ range and the restricted invariant mass of 
the jets lead to a reduction of the 2-jet rate at low $\xi$ and a rise 
at large $\xi$ with increasing $Q^2$.
	
\subsection{Extraction of the strong coupling}\label{sect:Extract}

The QCD predictions for jet production depend on $\alps$ and on the
parton density functions of the proton. The strong coupling $\alpha_s$
is determined from the measured normalised jet cross sections using
the parton density functions from global analyses, which include
inclusive deep-inelastic scattering and other data. The determination
is performed from individual observables and also from their combination.

QCD predictions of the jet cross sections are calculated as a function of $\alpha_s(\mu_r)$ with the FastNLO package using the CTEQ6.5M proton PDFs and applying the hadronisation corrections as described in section~\ref{sec:nlo}.
Measurements and theory predictions are used to calculate a $\chi^2(\alps)$ with the Hessian method \cite{Botje:1999dj}, where parameters representing systematic shifts of detector related observables are left free in the fit. The shifts in the electron energy scale, electron polar angle and the hadronic final state energy scale found by the fit are consistent with the \textit{a priori} estimated uncertainties. This method takes into account correlations of experimental uncertainties  and has also been used in global data analyses~\cite{Botje:1999dj,Barone:1999yv} and in previous H1 publications~\cite{H1Incl:2007pb, Adloff:2000dp}. The experimental uncertainty of $\alps$ is defined by the change in $\alps$ which gives an increase in $\chi^2$ of one unit with respect to the minimal value. 

The correlations of the experimental uncertainties between data points were estimated using Monte Carlo simulations:
\begin{itemize}
\item The statistical correlations between different observables using
  the same events are taken into account via the correlation matrix
  given in tables \ref{tMatrix13} and \ref{tMatrix46}. 
\item It is estimated that the uncertainty of the LAr hadronic energy scale is equally shared between correlated and uncorrelated contributions \cite{H1Incl:2007pb, Gouzevitch:2008zz}, while that from the electron energy scale is estimated to be $3/4$ uncorrelated \cite{Adloff:2003uh}.  
\item The measurement of the electron polar angle is assumed to be fully correlated~\cite{Adloff:2003uh}.
\item The model dependence of the experimental correction factors is considered as fully uncorrelated after the averaging procedure described in section \ref{sUncert}.
\end{itemize}
The sharing of correlated and uncorrelated contributions between the
different sources of uncertainty has the following impact on the
$\alps$ determination: when going from uncorrelated to fully
correlated error for each source, the fitted value of $\alps$
typically varies by half the total
experimental error and the estimated uncertainty by
less than $0.1\%$ of $\alps$.

The theory error is estimated by the so called \textit{offset} method
as the difference between the value of $\alps$ from the nominal fit to
the value when the fit is repeated with independent variations of
different sources of theoretical uncertainties as described in section
\ref{sec:nlo}. The resulting uncertainties due to the different
sources are summed in quadrature. The up (or down) variations are
applied simultaneously to all bins in the fit. The impact of
hadronisation corrections on $\alps$ is between $0.4\%$ and $1.0\%$,
while that of the factorisation scale amounts to $0.5\%$. The
sensitivity of $\alps$ to the renormalisation scale variation of the
inclusive NC DIS cross section alone is typically $0.5\%$. The largest
uncertainty, of typically $3\%$ to $4\%$, corresponds to the accuracy of
the NLO approximation to the jet cross sections estimated by varying
the renormalisation scale as described in section \ref{sec:nlo}. An
alternative method to estimate the impact of missing orders, called
the band method, developed by Jones et al.~\cite{Jones:2003yv} was
tested and, for the present measurement, it leads to a smaller
uncertainty on $\alps$ of typically $2\%$.

The uncertainty due to PDFs is estimated by propagating the CTEQ6.5M
errors. The typical size of the resulting error is $1.5\%$ for $\alps$
determined from the normalised inclusive jet or 2-jet cross sections
and $0.8\%$ when measured with the normalised 3-jet cross 
sections. This uncertainty is twice as large as that estimated with 
the uncertainties given for the MSTW2008nlo90cl set 
\cite{Martin:2009iq} which in turn exceeds the difference between 
$\alps$ values extracted with the central sets of CTEQ6.5M and 
MSTW2008nlo. The PDFs also depend on the value of $\alps$. Potential 
biases on the $\alps$ extraction from that source have been studied 
in detail previously~\cite{H1Incl:2007pb}. 
For this analysis, the resulting uncertainty is found to be 
negligible.

\begin{figure}[p]
\begin{minipage}{18pc}
\begin{flushleft}
\epsfig{file=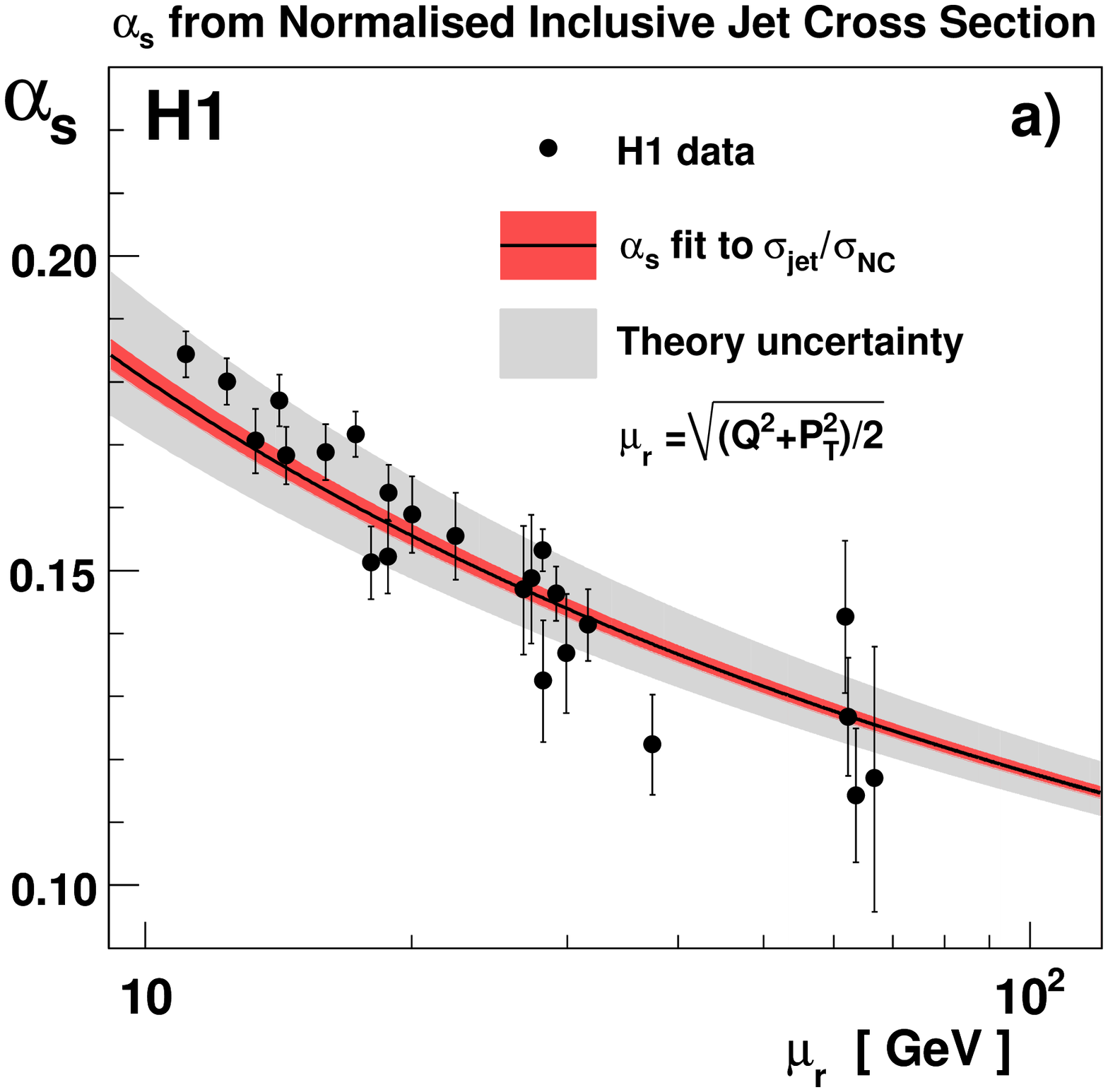,
width=77mm, angle=0, clip= }
\end{flushleft}
\end{minipage}
\begin{minipage}{18pc}
\begin{flushleft}
\epsfig{file=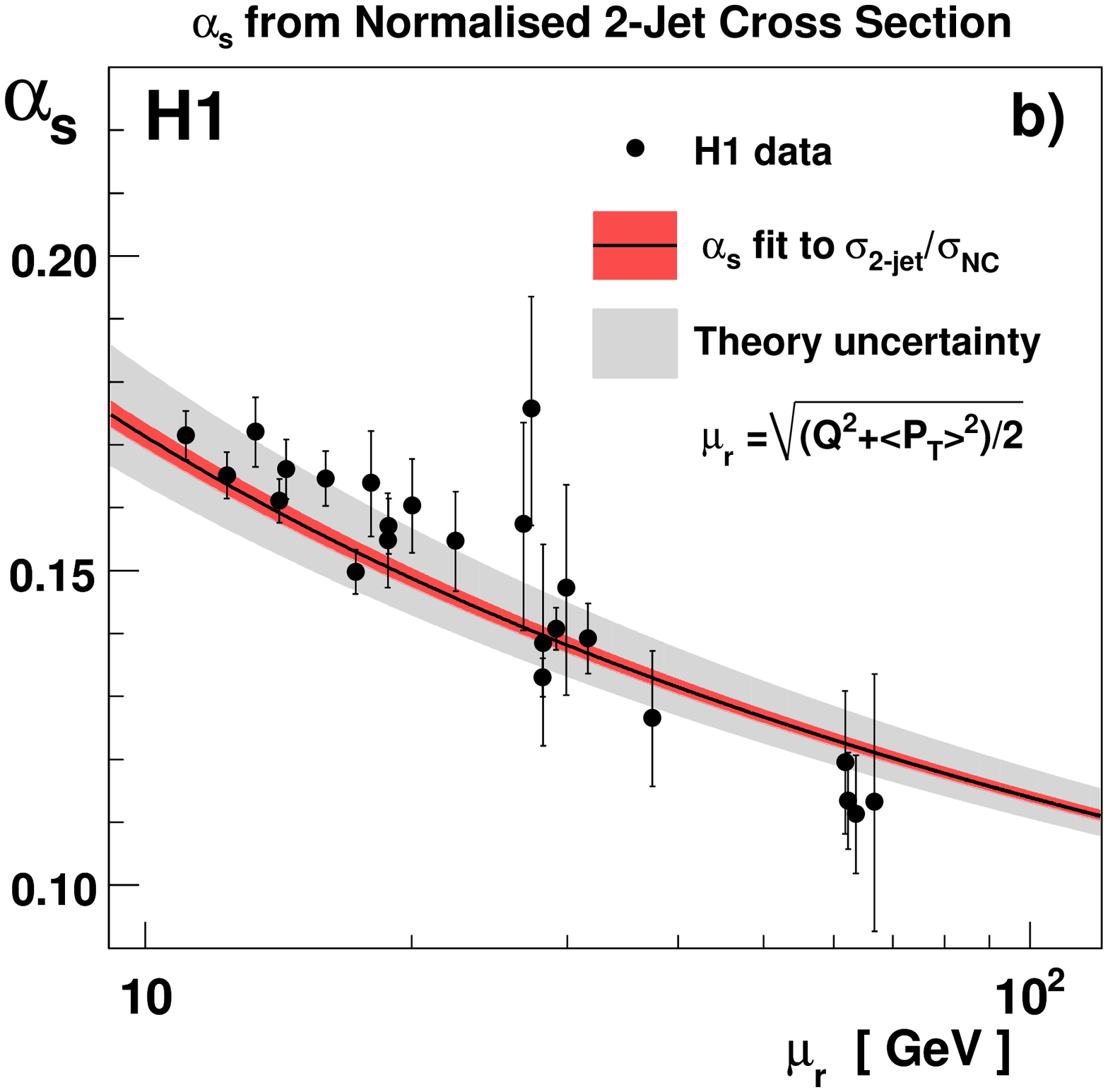,
width=77mm,angle=0,clip= }
\end{flushleft}
\end{minipage}

\caption{\label{fig::FitIjet1D_MuR} The $\alpha_s(\mu_{r}=\sqrt{(Q^2+P_T^2)/2})$ values determined using the
normalised inclusive jet cross sections (a) and the 2-jet cross sections (b), each measured in 24 bins of $Q^2$ and $P_T$. The error bars denote the total experimental uncertainty of 
each data point. 
In each plot, the solid line shows the two loop solution of the renormalisation group equation, resulting from evolving the $\alpha_s(M_Z)$ 
obtained from a simultaneous fit of all 24 measurements. The inner band denotes the experimental uncertainties and 
the outer band denotes the theoretical uncertainties associated with the renormalisation
and factorisation scales, the PDF uncertainty and the model dependence of the hadronisation 
corrections.}

\end{figure}

\begin{figure}[p]
\centering
\epsfig{file=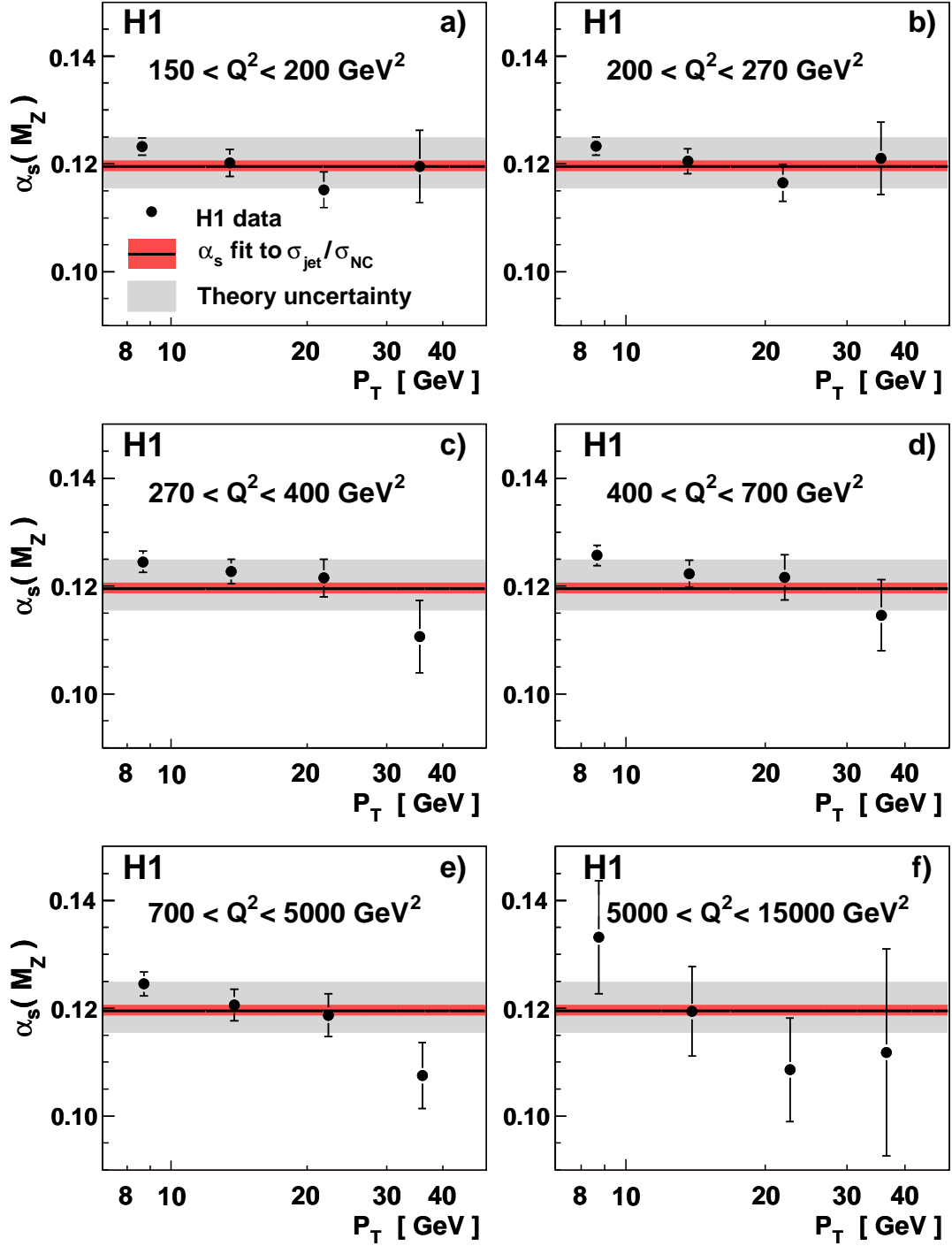,
width=148mm,angle=0,clip= }

\caption{\label{fig::FitIjet2D} The $\alpha_s(M_Z)$ values determined using the
normalised inclusive jet cross sections measured in 24 bins in $Q^2$ and $P_T$. Other details are given in the caption to figure \ref{fig::FitIjet1D_MuR}.}

\end{figure}

Individual fits of $\alpha_s(\mu_r)$ are made to each of the 24
measurements of the normalised double differential inclusive jet cross
section, as shown in figure \ref{fig::FitIjet1D_MuR}a.
These individual determinations show the
expected scale dependence. Equivalently, the $\alpha_s$ values at each
scale $\mu_r$ can be related to the value of the strong coupling
$\alpha_s(M_Z)$ at the $Z^0$ mass as shown in
figure~\ref{fig::FitIjet2D}.

\begin{figure}[p]
\begin{center}
\begin{minipage}{18pc}
\begin{flushleft}
\epsfig{file=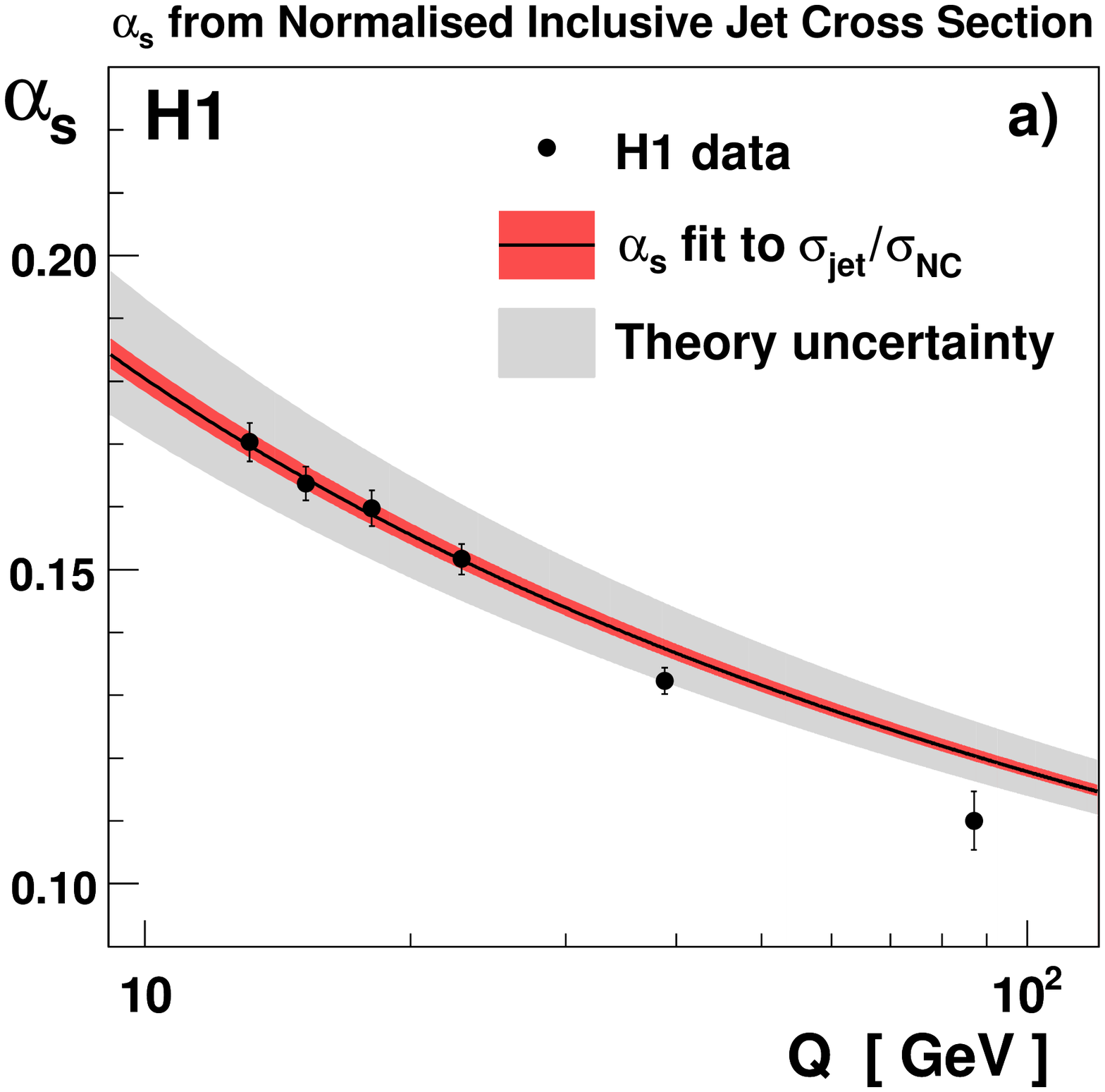,
width=77mm,angle=0,clip= }
\end{flushleft}
\end{minipage}
\begin{minipage}{18pc}
\begin{flushright}
\epsfig{file=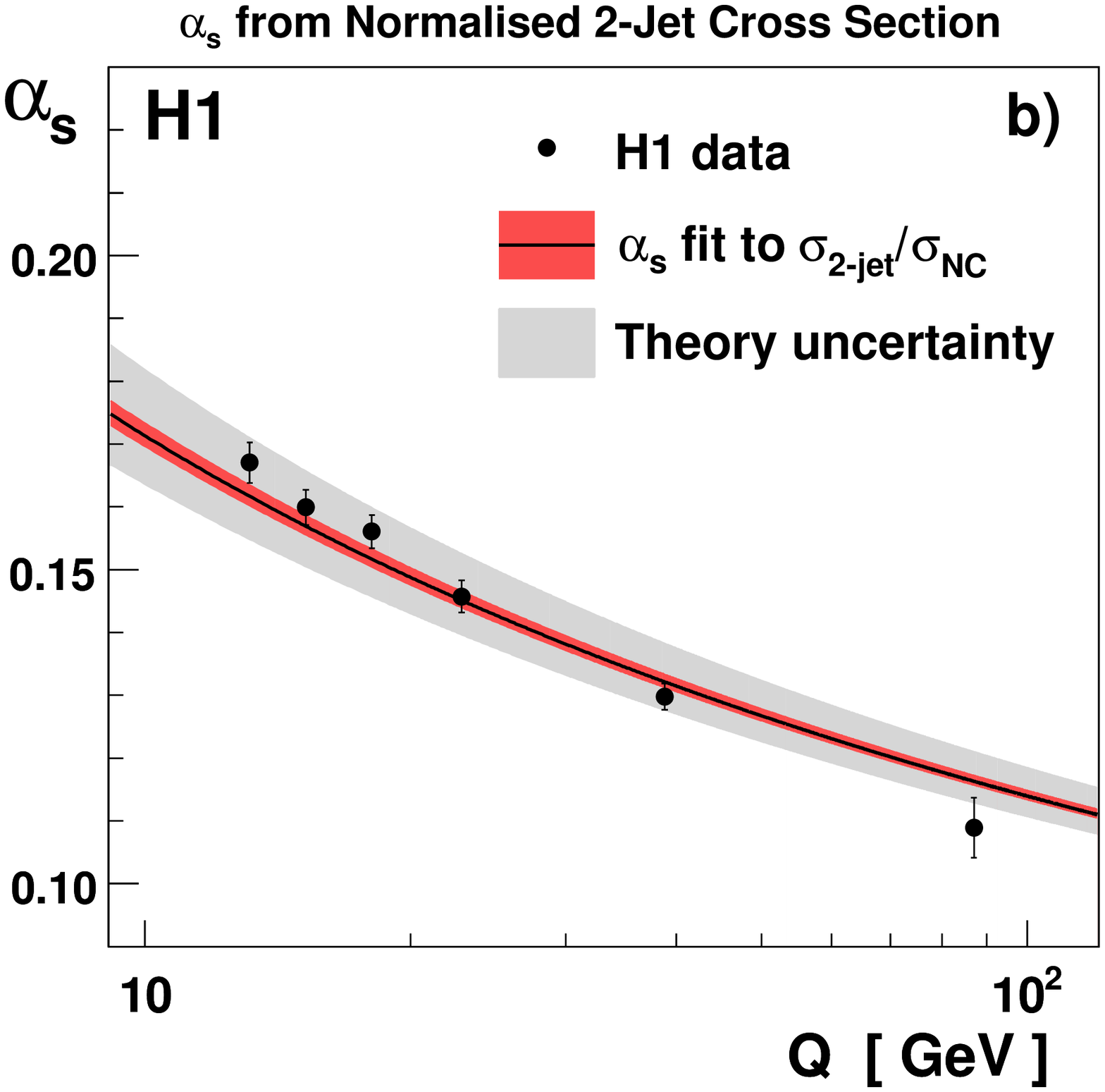,
width=77mm,angle=0,clip= }
\end{flushright}
\end{minipage}
\end{center}
\centering

\epsfig{file=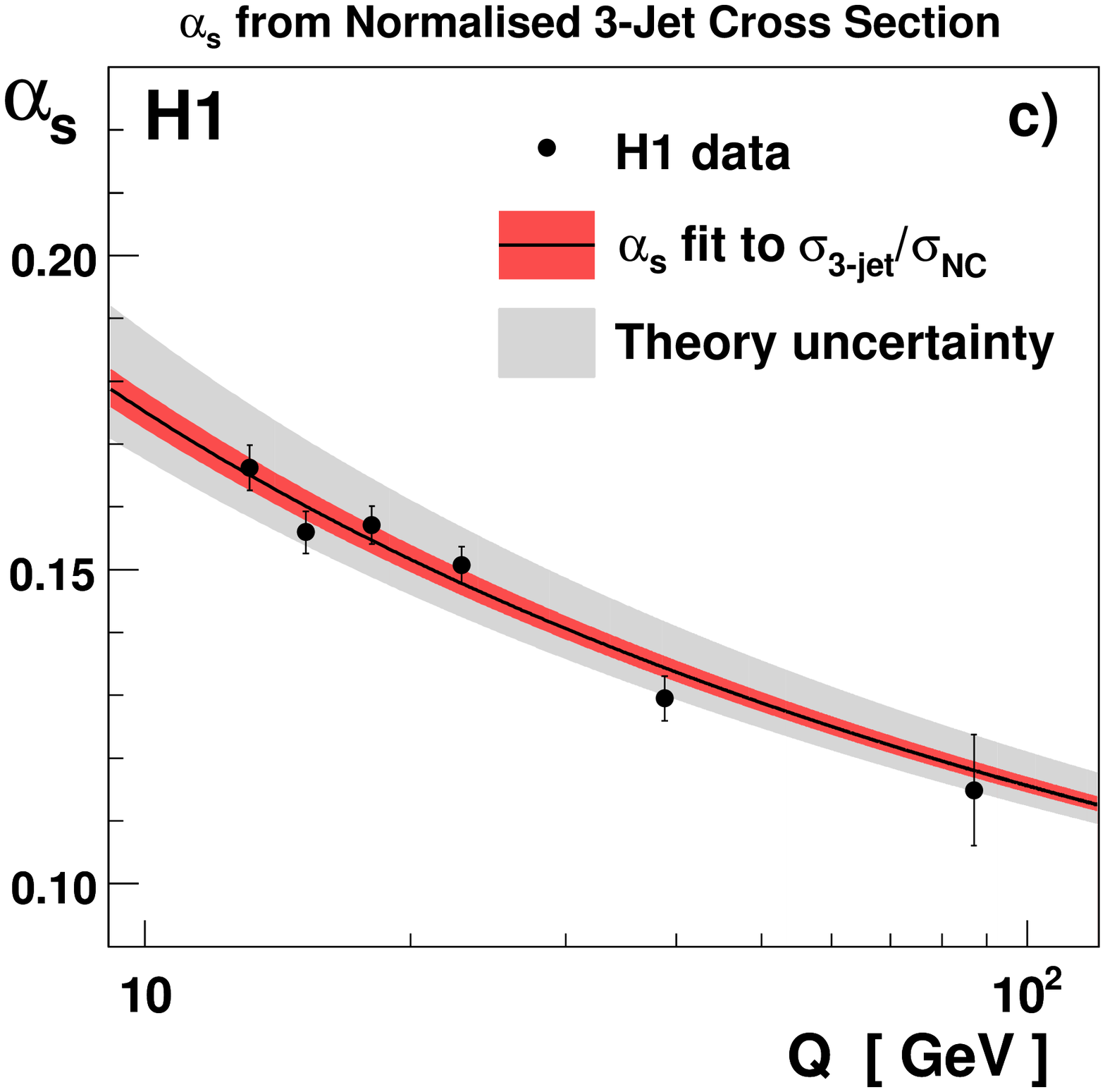,
width=75mm,angle=0,clip= }

\caption{\label{fig::Fitjet1D} The $\alpha_s(Q)$ values extracted by fitting the $P_T$ dependence of the normalised inclusive jet cross section in different regions of $Q^2$ (a); $\alpha_s(Q)$ values extracted by fitting the $\left\langle P_{T}\right\rangle$ dependence of the normalised 2-jet cross section in different regions of $Q^2$ (b); $\alpha_s(Q)$ values extracted from the normalised 3-jet cross section in different regions of $Q^2$ (c). In each case, the solid lines shows the two loop solution of the renormalisation group equation obtained by evolving the corresponding central value of the $\alpha_s(M_Z)$. Other details are given in the caption to \mbox{figure \ref{fig::FitIjet1D_MuR}}.}
\end{figure}

Then $\alpha_s(M_Z)$ is determined by a common fit to the normalised inclusive 
jet cross section in four $P_T$ bins for each region in $Q^2$. The resulting 
six values are evolved from the scale $M_Z$ to the average $Q$ in that region 
(figure~\ref{fig::Fitjet1D}a). Finally, a central value $\alpha_s(M_Z)$ is 
extracted from a common fit to all 24 measurements and given in 
table \ref{tab::Fits}. The result of evolving this value together with its 
associated uncertainty is also shown as the curve and surrounding band in 
figure \ref{fig::Fitjet1D}a.

\begin{figure}[p]
\begin{center}
\epsfig{file=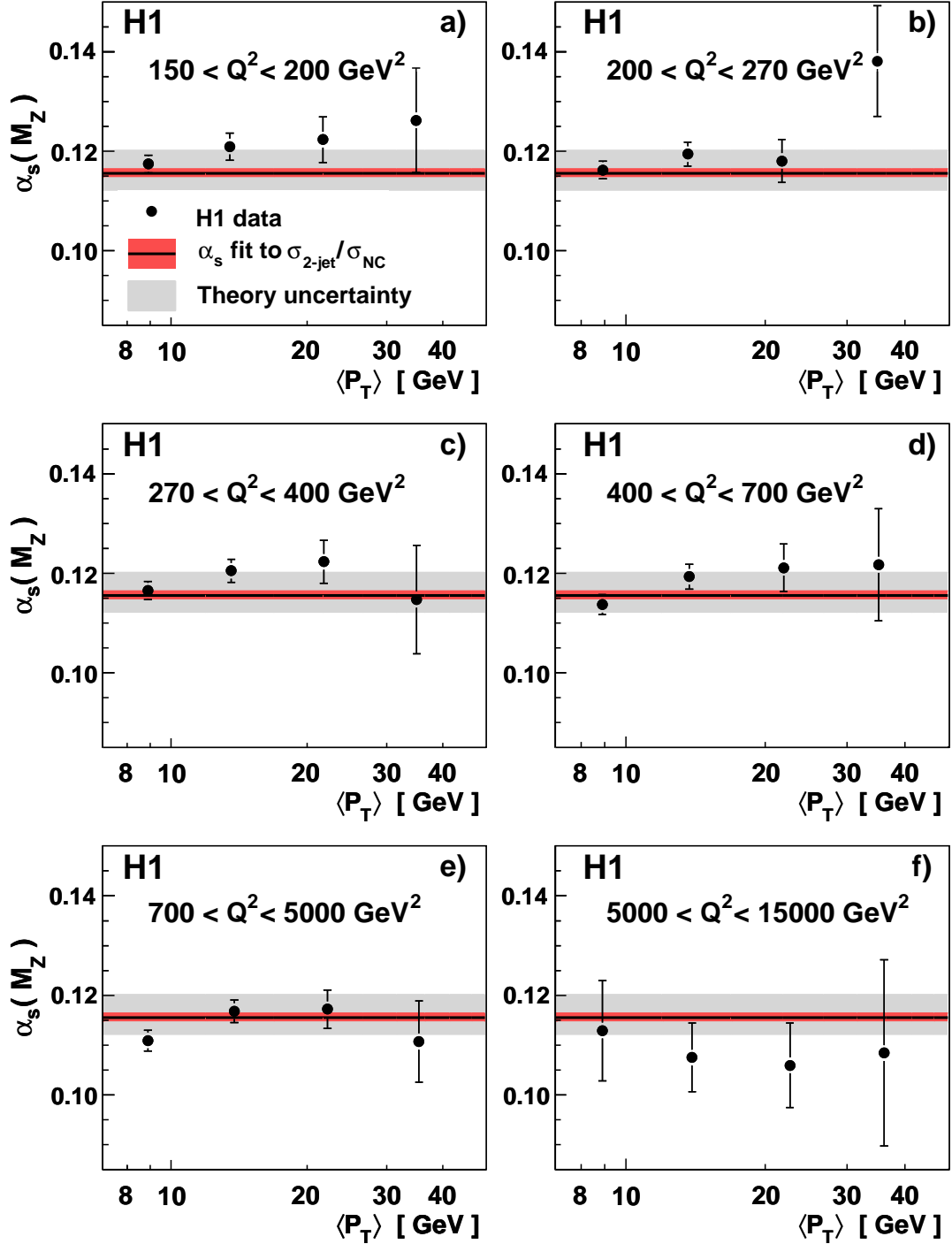,
width=148mm,angle=0,clip= }
\end{center}
\caption{\label{fig::Fit2jet2D} The $\alpha_s(M_Z)$ values determined using the
normalised 2-jet cross sections measured in 24 bins in $Q^2$ and $\left\langle P_{T}\right\rangle$. The solid line shows the two loop solution of the renormalisation group equation, $\alpha_s(M_Z)$, obtained from a simultaneous fit of all 24 measurements of the normalised 2-jet cross sections. Other details are given in the caption to figure \ref{fig::FitIjet1D_MuR}.}
\end{figure}

	The same fit procedure of successive combination steps is applied to the 24 points of the normalised 2-jet cross section with $\left\langle P_T \right\rangle > 7~$GeV  (figure \ref{fig::FitIjet1D_MuR}b, \ref{fig::Fit2jet2D} and \ref{fig::Fitjet1D}b). The bins with $5<\left\langle P_T \right\rangle < 7~$GeV are not used for the extraction of the strong coupling since the theory uncertainty is significantly larger than in the other bins (figure \ref{fig::2jet_Q2ET}). The fit procedure is also applied to the 6 points of the normalised 3-jet cross section (figure \ref{fig::Fitjet1D}c).
The normalised 3-jet cross section (figure \ref{fig::jet_Q2}c), which is $\mathcal{O}(\alpha_s^2)$, is preferred to the ratio of the 3-jet cross section to the 2-jet cross section (figure \ref{fig::jet_Q2}d), which is $\mathcal{O}(\alpha_s^1)$, due to better sensitivity to the strong coupling. The three values of $\alpha_s(M_Z)$ determined from the normalised inclusive jet (24 points), 2-jet (24 points) and 3-jet (6 points) cross sections are given in table \ref{tab::Fits} with experimental and theoretical uncertainties. All obtained values are compatible with each other within two standard deviations of the experimental uncertainty. 

\begin{figure}[p]
\centering
\epsfig{file=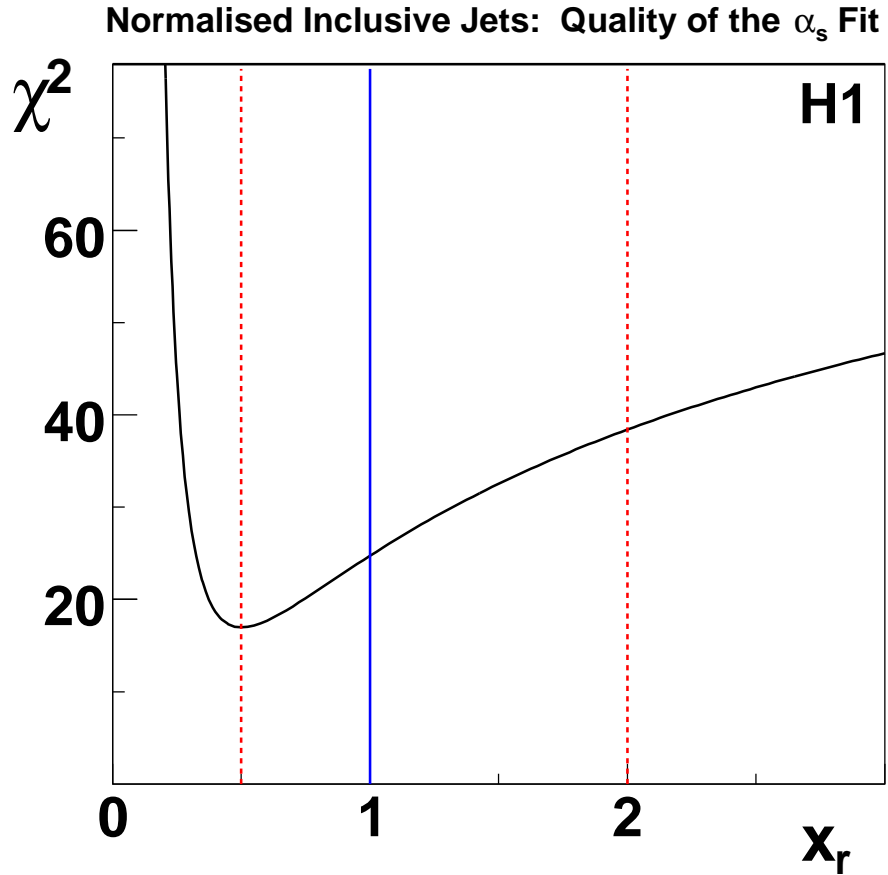,
width=130mm,angle=0,clip= }

\caption{\label{fig::FitChi2_MuR} The minimal $\chi^2$ of the fit of
  the NLO prediction with $\mu_r = x_r\cdot \sqrt{(Q^2+P_T^{2})/2}$ to
  the normalised inclusive jet cross section as function of $x_r$ for 
  23 degrees of freedom. Vertical dashed lines represent the range 
  where the renormalisation scale is varied in order to estimate the 
  impact of missing orders beyond NLO, while the full line indicates 
  the nominal choice of the scale with $x_r=1$.}

\end{figure}

The impact of the choice of renormalisation scale on the central value
of $\alpha_s(M_Z)$ is studied in the case of the normalised inclusive
jet cross section by repeating the fit procedure with \mbox{$\mu_r =
  P_T$} and $\mu_r = Q$ instead of $\mu_r =
  \sqrt{(Q^2+P_T^{2})/2}$. In the first case the central value of the
$\alpha_s(M_Z)$ is found to be approximatively $0.7\%$ smaller and in
the latter approximatively $1.5\%$ bigger with respect to the nominal
fit, a difference which is well inside the estimated theoretical
uncertainties. Similar deviations are observed for the normalised
2-jet and 3-jet cross sections when $\mu_r=Q$ is used instead of
$\mu_r = \sqrt{(Q^2+P_{T\raisebox{0pt}[0pt][0pt]{,\,}{\rm obs}}^{2})/2}$. To get information
on the description of the data by the NLO calculations as a function
of the renormalisation scale, the $\chi^2$ of the fit is studied in
the case of the normalised inclusive jet cross section for different
values of the parameter $x_r$, defined by $\mu_r=x_r \cdot
\sqrt{(Q^2+P_T^{2})/2}$. 
The results are shown in figure \ref{fig::FitChi2_MuR}, where the $\alps$
fit is repeated for different choices of $x_r$ and the corresponding
$\chi^2$ values are shown.
The lowest $\chi^2$ value is obtained for $x_r\simeq 0.5$ while $x_r$ choices above $2.0$ and \mbox{below $0.3$} are disfavoured. 

The sensitivity of the strong coupling determination procedure to the choice of the jet definition is tested for the normalised inclusive jet and 2-jet cross sections by repeating all the extraction procedure using the anti-$k_T$ metric \cite{Cacciari:2008gp} instead of $k_T$, but keeping the recombination scheme and the distance parameter unchanged. The resulting central value of $\alpha_s(M_Z)$ differs in both cases by less than $0.6\%$ from the central value extracted using the $k_T$ metric.

\begin{figure}[p]
\begin{center}
\epsfig{file=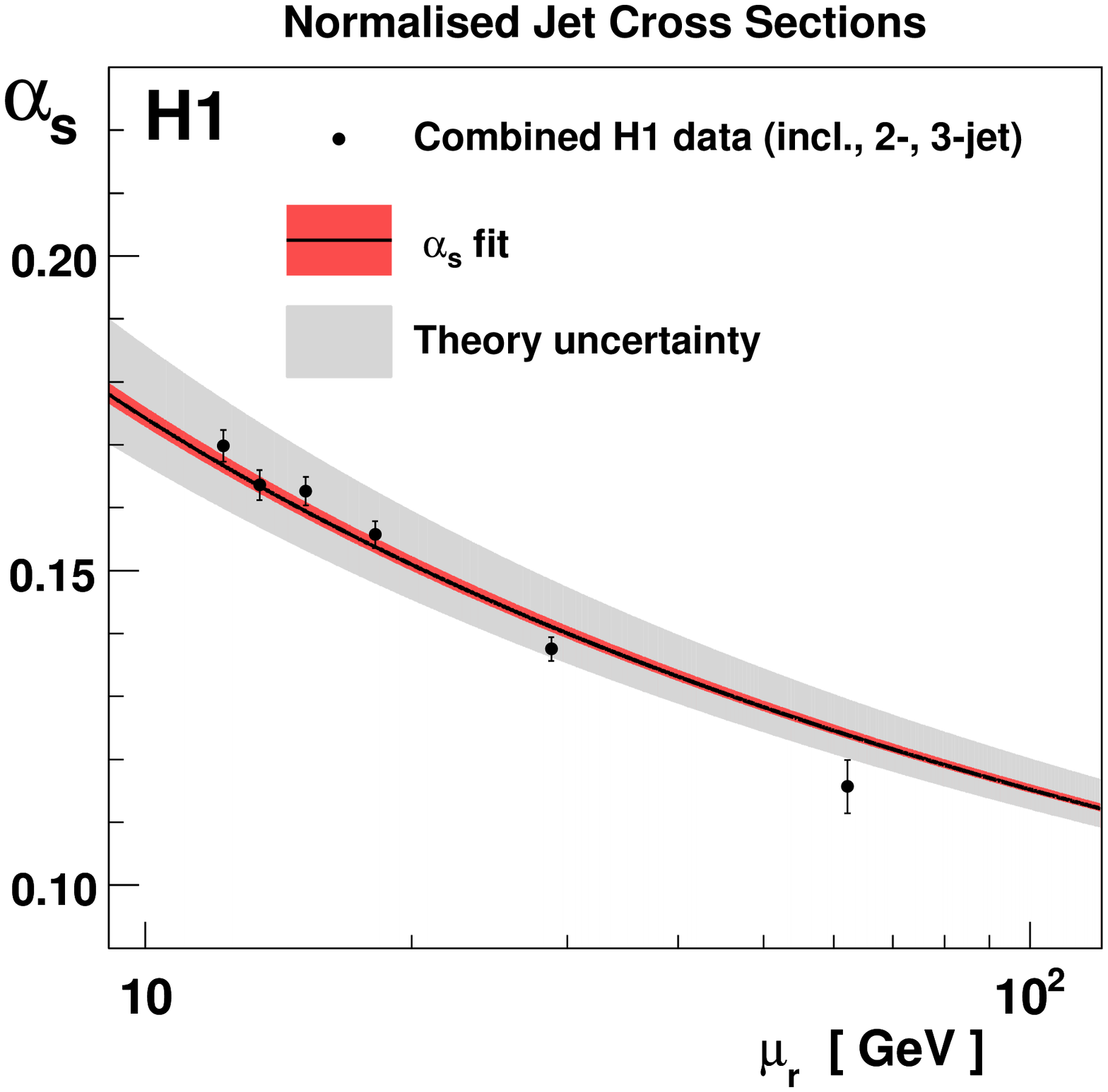,
width=130mm,angle=0,clip= }
\end{center}
\caption{\label{fig::FitAlljet1D} The values of $\alpha_s(\mu_r)$ obtained by a simultaneous fit of all normalised jet cross sections in each $Q^2$ bin. The solid line shows the two loop solution of the renormalisation group equation obtained by evolving the $\alpha_s$ extracted
from a simultaneous fit of 54 measurements of the normalised inclusive jet cross section as a function of $Q^2$ and $P_T$,
the normalised 2-jet cross section as function of $Q^2$ and $\left\langle P_{T}\right\rangle$ and the normalised 3-jet cross section as function of $Q^2$. Other details are given in the caption to figure \ref{fig::FitIjet1D_MuR}.}
\end{figure}

In each $Q^2$ region the values of $\alpha_s(M_Z)$ from
different observables are combined taking into account
statistical and systematic correlations. The resulting values,
evolved from the scale $M_Z$ to the average $\mu_r$ of the measurements in each
$Q^2$ region, are shown in figure \ref{fig::FitAlljet1D}. This
visualises the running of $\alps$ for scales 
between 10 and 100 GeV and the corresponding experimental and theory uncertainties. 
All 54 data points are used in a common fit of the strong coupling taking the correlations into account with a fit quality $\chi^2/{\rm ndf} = 65.0/53$ 
(see table \ref{tab::Fits}), which is also shown in figure \ref{fig::FitAlljet1D}.

The values of $\alpha_s(M_Z)$ obtained in this way are also consistent with 
the world averages $\alpha_s(M_Z)=0.1176 \pm 0.0020$ ~\cite{Amsler:2008zzb} and 
$\alpha_s(M_Z)=0.1189 \pm 0.0010$~\cite{Bethke:2006ac}, and with the previous 
H1 and ZEUS determinations from inclusive jet production 
measurements~\cite{H1Incl:2007pb, Chekanov:2006yc} and multijet 
production~\cite{Chekanov:2005ve}. 
The experimental error on $\alpha_s(M_Z)$ measured with each observable typically amounts to $1\%$. The combination of different observables, even though partially correlated, gives rise to additional constraints on the strong coupling and leads to an improved experimental uncertainty of $0.6\%$. The experimental error on $\alps$ is independent of the
choice of renormalisation scale within the variation used to determine the 
theoretical uncertainty. The total error is strongly dominated by the 
theoretical uncertainty due to missing higher orders in the 
perturbative calculation which is about $4\%$.

\section{Conclusion}

Measurements of the normalised inclusive, 2-jet and 3-jet cross sections in the Breit frame in deep-inelastic electron-proton scattering in the range $150<Q^2<15000\gev^2$ and $0.2<y<0.7$ using the H1 data taken in years 1999 to 2007 are presented.
Calculations at NLO QCD, corrected for hadronisation effects, provide
a good description of the single and double differential cross
sections as functions of the jet transverse momentum $P_T$, the boson
virtuality $Q^2$ as well as of the proton momentum fraction $\xi$. The
strong coupling $\alpha_s$ is determined from a fit of the NLO
prediction to the measured normalised jet cross sections. The
normalisation leads to cancellations of systematic effects,
resulting in improved experimental and PDF uncertainties. The
experimentally most precise determination of $\alpha_s(M_Z)$ is
derived from a common fit to the normalised jet cross sections:
\begin{eqnarray}
\nonumber
\alpha_s(M_Z) = 0.1168 ~\pm 0.0007 \,\mathrm{(exp.)}
~ ^{+0.0046}_{-0.0030}\,\mathrm{(th.)}~ \pm 0.0016\,(\text{\scshape pdf})\,. 
\end{eqnarray}
\noindent
The dominating source of the uncertainty is due to the renormalisation
scale dependence, which is used to estimate the effect of missing
higher orders beyond NLO in the pQCD prediction. This measurement
improves the experimental precision on $\alps$ determinations from
other recent jet measurements at HERA~\cite{H1Incl:2007pb,
Chekanov:2006yc}. The result is competitive with those from $e^+ 
e^-$ data~\cite{Bethke:2006ac, Dissertori:2007xa} and is in good 
agreement with the world average~\cite{Amsler:2008zzb,Bethke:2006ac}.

\section*{Acknowledgements}

We are grateful to the HERA machine group whose outstanding efforts have made this experiment possible. We thank the engineers and technicians for their work in constructing and maintaining the H1 detector, our funding agencies for financial support, the DESY technical staff for continual assistance and the DESY directorate for support and for the hospitality which they extend to the non DESY members of the collaboration. Furthermore we thank Gavin Salam, Matteo Cacciari, Mrinal Dasgupta and Zoltan Nagy for fruitful discussions.

\begin{table}[p]
	\centering
	\renewcommand{\arraystretch}{1.50}
  \begin{tabular}{| r | r | c | c |}
			\hline 
%
	NC DIS Selection & \multicolumn{3}{|c|}{$150 < Q^2 < 15000$~GeV$^2$ \ \
	\ \ \ \ \ $0.2 < y < 0.7$}  \\ \hline

     	Inclusive jet & $7 < P_{T} < 50\gev$                              & &  \multirow{3}{*}{$-0.8 < \eta_{\rm Lab}^{\rm jet} < 2.0$}  \\ \cline{1-3}      	  2-jet         & $5 < P_{T}^{\rm jet1},~P_{T}^{\rm jet2} < 50\gev$ & \multirow{2}{*}{$M_{12}>16\gev$}  & \\ \cline{1-2}
     	3-jet         & $5 < P_{T}^{\rm jet1},~P_{T}^{\rm jet2},~P_{T}^{\rm jet3} < 50\gev$ &   &  \\ \hline

  \end{tabular}
   \vspace{\baselineskip} 
	\caption{\label{tab::PhaseSpace} Selection criteria for the NC DIS and jet samples.}
\end{table}

\begin{table}[p]
	\footnotesize
		\renewcommand{\arraystretch}{1.50}

		\centering

	  \begin{tabular}{c r} \hline
		bin number & corresponding $Q^2$ range \\ \hline
 		1 & $150\le Q^2 < ~~~~200~\gevsq$ \\  
 		2 & $200\le Q^2 < ~~~~270~\gevsq$ \\  
 		3 & $270\le Q^2 < ~~~~400~\gevsq$ \\  
 		4 & $400\le Q^2 < ~~~~700~\gevsq$ \\  
 		5 & $700\le Q^2 < ~~5000~\gevsq$ \\ 
 		6 & $5000\le Q^2 < 15000~\gevsq$ \\   \hline
		\end{tabular}

 \vspace{\baselineskip} 
 \vspace{\baselineskip}

	\begin{minipage}{17pc}
		\centering
		\begin{tabular}{c c } \hline
		bin letter & corresponding $P_T$ or $\left\langle P_T\right\rangle$ range \\ \hline
 		~a' & $~~5\le P_T < ~~7~\gev$ \\  
 		a  &  $~~7\le P_T < 11~\gev$ \\  
 		b  &  $11\le P_T < 18~\gev$ \\  
 		c  &  $18\le P_T < 30~\gev$ \\  
 		d  &  $30\le P_T < 50~\gev$ \\  \hline
 		\end{tabular}

 	 \end{minipage}
	\begin{minipage}{17pc}
		\centering
 		\begin{tabular}{c c } \hline
		bin letter & corresponding $\xi$ range \\ \hline
		& \\  
 		A & $0.006\le \xi < 0.020$ \\  
 		B & $0.020\le \xi < 0.040$ \\  
 		C & $0.040\le \xi < 0.080$ \\  
 		D & $0.080\le \xi < 0.316$ \\
		 \hline  
  	\end{tabular}

  \end{minipage}
 	 \vspace{\baselineskip} 
	\caption{\label{tBinning} Nomenclature for the bins in $Q^2$, $P_T$ for the inclusive jet or $\left\langle P_T\right\rangle$ for 2-jets and $\xi$ used in the following tables. In case of the normalised 2-jet cross section, the bin $a'$ in $\left\langle P_T\right\rangle$ is not used for the $\alpha_s$ extraction.}
\end{table}

\begin{table}[htp]
\centering
\tiny \sf
\renewcommand{\arraystretch}{1.50}
\begin{tabular}{|c || r || r | r | r | r || r | r | r || r | r|} 

\multicolumn{11}{c}{ } \\
\multicolumn{11}{c}{\normalsize Normalised inclusive jet cross section in bins of $Q^{2}$}\\
\multicolumn{11}{c}{ } \\
\hline
 & & &  & total & total &
\multicolumn{3}{|c||}{\underline{\hspace*{0.1cm}single contributions to correlated uncertainty\hspace*{0.1cm}}
}
   &                                           hadronisation & hadronisation\\
bin
& normalised   & statistical  & total        & uncorrelated & correlated
& electron     & electron     & hadronic    & correction    & correction    \\

& cross        & uncert.      & uncert.  & uncertainty  & uncert.
& energy scale & polar angle  & energy scale &  factor    & uncertainty \\

& section      & (\%) & (\%) & (\%) & (\%)
& (\%) & (\%) & (\%) &       & (\%)\\
\hline
1 &$2.39\, 10^{-1}	$&$0.7	$&$3.2	$&$2.4	$&$2.1	$&$0.6	$&$0.8	$&$1.9	$&$0.95	$&$0.6$ \\ 
\hline 
2 &$2.69\, 10^{-1}	$&$0.7	$&$3.0	$&$2.2	$&$1.9	$&$0.6	$&$0.6	$&$1.7	$&$0.94	$&$0.6$ \\ 
\hline 
3 &$3.11\, 10^{-1}	$&$0.8	$&$2.9	$&$2.3	$&$1.8	$&$0.6	$&$0.4	$&$1.6	$&$0.94	$&$0.8$ \\ 
\hline 
4 &$3.62\, 10^{-1}	$&$0.8	$&$2.7	$&$2.2	$&$1.7	$&$0.7	$&$0.3	$&$1.5	$&$0.94	$&$0.6$ \\ 
\hline 
5 &$4.26\, 10^{-1}	$&$0.9	$&$2.7	$&$2.3	$&$1.6	$&$0.9	$&$0.1	$&$1.3	$&$0.93	$&$1.7$ \\ 
\hline 
6 &$5.02\, 10^{-1}	$&$3.2	$&$5.7	$&$5.2	$&$2.4	$&$2.2	$&$0.3	$&$0.8	$&$0.93	$&$3.0$ \\ 
\hline 

\multicolumn{11}{c}{ } \\
\multicolumn{11}{c}{\normalsize Normalised 2-jet cross section in bins of $Q^{2}$}\\
\multicolumn{11}{c}{ } \\
\hline 
1 &$8.81\, 10^{-2}	$&$1.0	$&$2.9	$&$2.2	$&$1.9	$&$0.4	$&$0.7	$&$1.7	$&$0.94	$&$1.1$ \\ 
\hline 
2 &$1.01\, 10^{-1}	$&$1.1	$&$2.6	$&$2.1	$&$1.7	$&$0.3	$&$0.6	$&$1.5	$&$0.93	$&$1.3$ \\ 
\hline 
3 &$1.19\, 10^{-1}	$&$1.1	$&$2.4	$&$1.9	$&$1.5	$&$0.2	$&$0.4	$&$1.4	$&$0.93	$&$1.3$ \\ 
\hline 
4 &$1.41\, 10^{-1}	$&$1.2	$&$2.5	$&$2.0	$&$1.4	$&$0.3	$&$0.4	$&$1.3	$&$0.91	$&$1.1$ \\ 
\hline 
5 &$1.75\, 10^{-1}	$&$1.2	$&$2.4	$&$2.1	$&$1.2	$&$0.1	$&$0.2	$&$1.1	$&$0.91	$&$2.9$ \\ 
\hline 
6 &$1.97\, 10^{-1}	$&$4.4	$&$7.7	$&$7.6	$&$1.1	$&$0.3	$&$0.2	$&$1.0	$&$0.91	$&$2.9$ \\ 
\hline 

\multicolumn{11}{c}{ } \\
\multicolumn{11}{c}{\normalsize Normalised 3-jet cross section in bins of $Q^{2}$}\\
\multicolumn{11}{c}{ } \\
\hline
1 &$1.19\, 10^{-2}	$&$2.6	$&$5.1	$&$4.1	$&$3.1	$&$0.4	$&$1.3	$&$2.8	$&$0.85	$&$2.4$ \\ 
\hline 
2 &$1.29\, 10^{-2}	$&$2.8	$&$5.1	$&$4.2	$&$2.9	$&$0.3	$&$0.7	$&$2.8	$&$0.84	$&$1.7$ \\ 
\hline 
3 &$1.68\, 10^{-2}	$&$2.7	$&$4.6	$&$3.8	$&$2.6	$&$0.1	$&$0.9	$&$2.5	$&$0.83	$&$1.0$ \\ 
\hline 
4 &$2.06\, 10^{-2}	$&$2.9	$&$4.7	$&$4.0	$&$2.5	$&$0.3	$&$0.8	$&$2.4	$&$0.82	$&$0.6$ \\ 
\hline 
5 &$2.36\, 10^{-2}	$&$2.8	$&$6.6	$&$6.2	$&$2.3	$&$0.4	$&$0.4	$&$2.2	$&$0.81	$&$1.2$ \\ 
\hline 
6 &$2.82\, 10^{-2}	$&$9.2	$&$18.7	$&$18.5	$&$2.3	$&$0.4	$&$0.3	$&$2.3	$&$0.75	$&$3.6$ \\ 
\hline 

\multicolumn{11}{c}{ } \\
\multicolumn{11}{c}{\normalsize 3-jet cross section normalised to 2-jet cross section in bins of $Q^{2}$}\\
\multicolumn{11}{c}{ } \\
\hline
1 &$1.36\, 10^{-1}	$&$2.7	$&$4.4	$&$4.3	$&$1.2	$&$0.2	$&$0.5	$&$1.1	$&$0.91	$&$1.5$ \\ 
\hline 
2 &$1.28\, 10^{-1}	$&$3.0	$&$4.7	$&$4.5	$&$1.4	$&$0.5	$&$0.1	$&$1.3	$&$0.90	$&$1.0$ \\ 
\hline 
3 &$1.41\, 10^{-1}	$&$2.9	$&$4.5	$&$4.3	$&$1.1	$&$0.2	$&$0.1	$&$1.1	$&$0.90	$&$0.9$ \\ 
\hline 
4 &$1.46\, 10^{-1}	$&$3.1	$&$4.8	$&$4.6	$&$1.2	$&$0.7	$&$0.3	$&$1.0	$&$0.90	$&$0.6$ \\ 
\hline 
5 &$1.35\, 10^{-1}	$&$3.0	$&$5.1	$&$5.0	$&$1.2	$&$0.4	$&$0.2	$&$1.1	$&$0.89	$&$1.5$ \\ 
\hline 
6 &$1.43\, 10^{-1}	$&$9.8	$&$14.2	$&$14.1	$&$1.3	$&$0.3	$&$0.2	$&$1.2	$&$0.82	$&$3.3$ \\ 
\hline 

\end{tabular} 
\normalfont
 \vspace{\baselineskip} 
\caption{\label{tab::jet_Q2} Normalised inclusive jet, 2-jet and 3-jet
  cross sections in NC DIS measured as a function of $Q^2$. The
  measurements refer to the phase-space defined in table
  \ref{tab::PhaseSpace}. In columns 3 to 9 are shown the 
  statistical uncertainty, the total experimental uncertainty, the 
  total uncorrelated uncertainty including the statistical one and the
  total correlated uncertainty calculated as the quadratic sum of the
  following three components: the electron energy scale, the electron
  polar angle uncertainty and the hadron energy scale uncertainty. The
  sharing of the uncertainties between correlated and uncorrelated
  sources is described in detail in section \ref{sect:Extract}. The
  hadronisation correction factors applied to the NLO predictions and
  their uncertainties are shown in columns 10 and 11. The bin
  nomenclature of column 1 is defined in table \ref{tBinning}.} 
\end{table}

\begin{table}[htp]
\tiny \sf
\centering
\renewcommand{\arraystretch}{1.50}
\begin{tabular}{|c || r || r | r | r | r || r | r | r || r | r|} 

\multicolumn{11}{c}{ } \\
\multicolumn{11}{c}{\normalsize Normalised inclusive jet cross section in bins of $Q^{2}$ and $P_{T}$}\\
\multicolumn{11}{c}{ } \\
\hline
 & & &  & total & total &
\multicolumn{3}{|c||}{\underline{\hspace*{0.1cm}single contributions to correlated uncertainty\hspace*{0.1cm}}
}
   &                                           hadronisation & hadronisation\\
bin
& normalised   & statistical  & total        & uncorrelated & correlated
& electron     & electron     & hadronic    & correction    & correction    \\

& cross        & uncert.      & uncert.  & uncertainty  & uncert.
& energy scale & polar angle  & energy scale &  factor    & uncertainty \\

& section      & (\%) & (\%) & (\%) & (\%)
& (\%) & (\%) & (\%) &       & (\%)\\
\hline
1 a&$1.53\, 10^{-1}	$&$0.8	$&$2.7	$&$2.1	$&$1.7	$&$0.6	$&$0.6	$&$1.4	$&$0.94	$&$0.7$ \\ 
1 b&$6.93\, 10^{-2}	$&$1.2	$&$4.5	$&$3.5	$&$2.9	$&$0.6	$&$1.1	$&$2.6	$&$0.97	$&$0.3$ \\ 
1 c&$1.53\, 10^{-2}	$&$2.5	$&$6.1	$&$4.7	$&$3.9	$&$0.6	$&$1.6	$&$3.5	$&$0.96	$&$0.6$ \\ 
1 d&$1.93\, 10^{-3}	$&$7.2	$&$10.6	$&$9.7	$&$4.4	$&$0.2	$&$1.3	$&$4.2	$&$0.95	$&$1.8$ \\ 
\hline 
2 a&$1.66\, 10^{-1}	$&$0.9	$&$2.6	$&$2.2	$&$1.4	$&$0.7	$&$0.4	$&$1.2	$&$0.93	$&$0.6$ \\ 
2 b&$8.10\, 10^{-2}	$&$1.3	$&$3.9	$&$3.0	$&$2.5	$&$0.6	$&$0.8	$&$2.3	$&$0.97	$&$0.4$ \\ 
2 c&$1.97\, 10^{-2}	$&$2.6	$&$5.8	$&$4.6	$&$3.6	$&$0.3	$&$0.9	$&$3.5	$&$0.96	$&$0.9$ \\ 
2 d&$2.67\, 10^{-3}	$&$7.1	$&$10.2	$&$9.1	$&$4.6	$&$0.4	$&$0.5	$&$4.5	$&$0.97	$&$3.2$ \\ 
\hline 
3 a&$1.82\, 10^{-1}	$&$1.0	$&$2.8	$&$2.4	$&$1.4	$&$0.7	$&$0.4	$&$1.1	$&$0.92	$&$0.7$ \\ 
3 b&$9.82\, 10^{-2}	$&$1.3	$&$3.5	$&$2.7	$&$2.2	$&$0.5	$&$0.4	$&$2.1	$&$0.97	$&$1.0$ \\ 
3 c&$2.76\, 10^{-2}	$&$2.4	$&$5.5	$&$4.4	$&$3.3	$&$0.3	$&$0.8	$&$3.2	$&$0.96	$&$0.4$ \\ 
3 d&$3.11\, 10^{-3}	$&$7.0	$&$9.8	$&$8.5	$&$4.8	$&$0.1	$&$1.9	$&$4.4	$&$0.95	$&$3.2$ \\ 
\hline 
4 a&$2.02\, 10^{-1}	$&$1.1	$&$2.4	$&$2.2	$&$1.1	$&$0.7	$&$0.2	$&$0.9	$&$0.92	$&$0.5$ \\ 
4 b&$1.16\, 10^{-1}	$&$1.3	$&$3.4	$&$2.8	$&$2.0	$&$0.8	$&$0.4	$&$1.8	$&$0.96	$&$0.5$ \\ 
4 c&$3.83\, 10^{-2}	$&$2.3	$&$5.9	$&$4.9	$&$3.3	$&$0.5	$&$0.7	$&$3.1	$&$0.97	$&$1.5$ \\ 
4 d&$5.28\, 10^{-3}	$&$6.3	$&$8.9	$&$7.9	$&$4.1	$&$0.3	$&$0.6	$&$4.1	$&$0.96	$&$2.7$ \\ 
\hline 
5 a&$2.13\, 10^{-1}	$&$1.2	$&$2.4	$&$2.1	$&$1.1	$&$0.8	$&$0.1	$&$0.7	$&$0.90	$&$2.4$ \\ 
5 b&$1.42\, 10^{-1}	$&$1.3	$&$3.3	$&$2.8	$&$1.7	$&$0.9	$&$0.1	$&$1.5	$&$0.96	$&$1.1$ \\ 
5 c&$5.91\, 10^{-2}	$&$2.0	$&$4.7	$&$3.8	$&$2.7	$&$0.9	$&$0.1	$&$2.6	$&$0.97	$&$0.3$ \\ 
5 d&$1.09\, 10^{-2}	$&$4.4	$&$7.4	$&$6.3	$&$3.9	$&$0.1	$&$0.3	$&$3.8	$&$0.96	$&$3.1$ \\ 
\hline 
6 a&$2.32\, 10^{-1}	$&$4.3	$&$8.1	$&$7.8	$&$2.3	$&$2.2	$&$0.4	$&$0.4	$&$0.90	$&$3.9$ \\ 
6 b&$1.62\, 10^{-1}	$&$4.8	$&$7.5	$&$6.9	$&$2.9	$&$2.8	$&$0.4	$&$0.7	$&$0.94	$&$2.5$ \\ 
6 c&$8.14\, 10^{-2}	$&$6.7	$&$9.8	$&$9.4	$&$2.6	$&$2.2	$&$0.1	$&$1.4	$&$0.96	$&$0.8$ \\ 
6 d&$2.66\, 10^{-2}	$&$9.7	$&$19.0	$&$18.8	$&$3.1	$&$0.6	$&$0.5	$&$3.0	$&$0.97	$&$3.6$ \\ 
\hline 
\end{tabular} 
\normalfont
 \vspace{\baselineskip} 
\caption{\label{tab::Ijet_Q2ET} Normalised inclusive jet cross sections as a function of $Q^2$ and $P_T$ together with their relative errors and hadronisation correction factors. Other details are given in the caption to table \ref{tab::jet_Q2}. The bin nomenclature is defined in table \ref{tBinning}.} 
\end{table}

\begin{table}[htp]
\tiny \sf
\centering
\renewcommand{\arraystretch}{1.50}
\begin{tabular}{|c || r || r | r | r | r || r | r | r || r | r|} 

\multicolumn{11}{c}{ } \\
\multicolumn{11}{c}{\normalsize Normalised 2-jet cross section in bins of $Q^{2}$ and $\left\langle P_{T} \right\rangle$}\\
\multicolumn{11}{c}{ } \\
\hline
 & & &  & total & total &
\multicolumn{3}{|c||}{\underline{\hspace*{0.1cm}single contributions to correlated uncertainty\hspace*{0.1cm}}
}
   &                                           hadronisation & hadronisation\\
bin
& normalised   & statistical  & total        & uncorrelated & correlated
& electron     & electron     & hadronic    & correction    & correction    \\

& cross        & uncert.      & uncert.  & uncertainty  & uncert.
& energy scale & polar angle  & energy scale &  factor    & uncertainty \\

& section      & (\%) & (\%) & (\%) & (\%)
& (\%) & (\%) & (\%) &       & (\%)\\
\hline
~1 a'&$9.14\, 10^{-3}	$&$3.2	$&$3.5	$&$3.5	$&$0.6	$&$0.5	$&$0.1	$&$0.3	$&$0.83	$&$2.5$ \\ 
1 a&$4.40\, 10^{-2}	$&$1.4	$&$2.9	$&$2.6	$&$1.4	$&$0.5	$&$0.5	$&$1.2	$&$0.94	$&$1.4$ \\ 
1 b&$2.77\, 10^{-2}	$&$1.7	$&$4.4	$&$3.6	$&$2.6	$&$0.4	$&$1.0	$&$2.3	$&$0.96	$&$1.4$ \\ 
1 c&$6.28\, 10^{-3}	$&$3.5	$&$6.8	$&$5.3	$&$4.2	$&$0.3	$&$1.9	$&$3.8	$&$0.96	$&$1.4$ \\ 
1 d&$6.87\, 10^{-4}	$&$10.5	$&$12.6	$&$11.7	$&$4.5	$&$0.1	$&$0.8	$&$4.5	$&$0.95	$&$1.8$ \\ 
\hline 
~2 a'&$1.04\, 10^{-2}	$&$3.5	$&$4.2	$&$4.1	$&$0.9	$&$0.2	$&$0.8	$&$0.2	$&$0.83	$&$1.1$ \\ 
2 a&$4.91\, 10^{-2}	$&$1.5	$&$2.8	$&$2.5	$&$1.2	$&$0.4	$&$0.5	$&$1.0	$&$0.94	$&$1.3$ \\ 
2 b&$3.26\, 10^{-2}	$&$1.8	$&$3.8	$&$3.0	$&$2.2	$&$0.4	$&$0.8	$&$2.1	$&$0.96	$&$1.9$ \\ 
2 c&$7.80\, 10^{-3}	$&$3.6	$&$6.3	$&$5.1	$&$3.6	$&$0.2	$&$1.1	$&$3.4	$&$0.96	$&$1.7$ \\ 
2 d&$1.05\, 10^{-3}	$&$10.1	$&$12.5	$&$11.5	$&$4.9	$&$0.4	$&$0.5	$&$4.8	$&$0.92	$&$3.3$ \\ 
\hline 
~3 a'&$1.13\, 10^{-2}	$&$3.7	$&$3.9	$&$3.9	$&$0.3	$&$0.3	$&$0.1	$&$0.2	$&$0.80	$&$1.4$ \\ 
3 a&$5.56\, 10^{-2}	$&$1.6	$&$2.6	$&$2.4	$&$0.9	$&$0.3	$&$0.3	$&$0.8	$&$0.92	$&$0.3$ \\ 
3 b&$3.99\, 10^{-2}	$&$1.8	$&$3.4	$&$2.9	$&$1.9	$&$0.4	$&$0.3	$&$1.8	$&$0.97	$&$2.2$ \\ 
3 c&$1.10\, 10^{-2}	$&$3.3	$&$5.9	$&$4.8	$&$3.3	$&$0.1	$&$0.8	$&$3.2	$&$0.96	$&$1.0$ \\ 
3 d&$1.16\, 10^{-3}	$&$10.1	$&$13.2	$&$11.8	$&$5.9	$&$0.2	$&$2.9	$&$5.2	$&$0.94	$&$2.7$ \\ 
\hline 
~4 a'&$1.41\, 10^{-2}	$&$3.9	$&$4.1	$&$4.1	$&$0.2	$&$0.1	$&$0.1	$&$0.2	$&$0.79	$&$2.7$ \\ 
4 a&$6.13\, 10^{-2}	$&$1.8	$&$2.7	$&$2.6	$&$0.8	$&$0.3	$&$0.1	$&$0.7	$&$0.90	$&$0.1$ \\ 
4 b&$4.80\, 10^{-2}	$&$1.9	$&$3.5	$&$2.9	$&$1.8	$&$0.6	$&$0.5	$&$1.7	$&$0.96	$&$1.3$ \\ 
4 c&$1.57\, 10^{-2}	$&$3.2	$&$6.3	$&$5.6	$&$3.0	$&$0.2	$&$0.6	$&$2.9	$&$0.97	$&$1.2$ \\ 
4 d&$2.09\, 10^{-3}	$&$9.1	$&$12.7	$&$11.8	$&$4.7	$&$0.1	$&$0.7	$&$4.7	$&$0.96	$&$2.6$ \\ 
\hline 
~5 a'&$1.53\, 10^{-2}	$&$4.2	$&$9.2	$&$9.2	$&$0.8	$&$0.7	$&$0.2	$&$0.0	$&$0.77	$&$2.3$ \\ 
5 a&$6.95\, 10^{-2}	$&$1.9	$&$2.6	$&$2.5	$&$0.5	$&$0.1	$&$0.2	$&$0.5	$&$0.89	$&$3.6$ \\ 
5 b&$5.98\, 10^{-2}	$&$1.9	$&$2.9	$&$2.6	$&$1.3	$&$0.3	$&$0.1	$&$1.3	$&$0.94	$&$1.1$ \\ 
5 c&$2.49\, 10^{-2}	$&$2.8	$&$4.7	$&$4.0	$&$2.5	$&$0.1	$&$0.5	$&$2.4	$&$0.97	$&$2.2$ \\ 
5 d&$4.34\, 10^{-3}	$&$6.4	$&$9.1	$&$8.1	$&$4.2	$&$0.4	$&$0.7	$&$4.1	$&$0.96	$&$2.5$ \\ 
\hline 
~6 a'&$1.30\, 10^{-2}	$&$16.2	$&$36.2	$&$36.1	$&$2.4	$&$2.3	$&$0.6	$&$0.6	$&$0.73	$&$16.2$ \\ 
6 a&$7.47\, 10^{-2}	$&$7.1	$&$10.6	$&$10.6	$&$0.5	$&$0.4	$&$0.1	$&$0.3	$&$0.88	$&$1.0$ \\ 
6 b&$6.42\, 10^{-2}	$&$6.9	$&$8.0	$&$7.8	$&$1.5	$&$0.9	$&$0.3	$&$1.1	$&$0.93	$&$4.4$ \\ 
6 c&$3.36\, 10^{-2}	$&$9.2	$&$9.7	$&$9.5	$&$1.6	$&$0.7	$&$0.4	$&$1.3	$&$0.95	$&$1.8$ \\ 
6 d&$1.03\, 10^{-2}	$&$15.7	$&$19.2	$&$18.8	$&$3.6	$&$0.6	$&$0.5	$&$3.5	$&$0.97	$&$4.3$ \\ 
\hline 
\end{tabular} 
\normalfont
 \vspace{\baselineskip} 
\caption{\label{tab::2jet_Q2ET} Normalised 2-jet cross sections as a function of $Q^2$ and $\left\langle  P_T \right\rangle$ together with their relative errors and hadronisation correction factors. Other details are given in the caption to table \ref{tab::jet_Q2}. The bin nomenclature is defined in table \ref{tBinning}.} 
\end{table}

\begin{table}[htp]
\tiny \sf
\centering
\renewcommand{\arraystretch}{1.50}
\begin{tabular}{|c || r || r | r | r | r || r | r | r || r | r|} 

\multicolumn{11}{c}{ } \\
\multicolumn{11}{c}{\normalsize Normalised 2-jet cross section in bins of $Q^{2}$ and $\xi$}\\
\multicolumn{11}{c}{ } \\
\hline
 & & &  & total & total &
\multicolumn{3}{|c||}{\underline{\hspace*{0.1cm}single contributions to correlated uncertainty\hspace*{0.1cm}}
}
   &                                           hadronisation & hadronisation\\
bin
& normalised   & statistical  & total        & uncorrelated & correlated
& electron     & electron     & hadronic    & correction    & correction    \\

& cross        & uncert.      & uncert.  & uncertainty  & uncert.
& energy scale & polar angle  & energy scale &  factor    & uncertainty \\

& section      & (\%) & (\%) & (\%) & (\%)
& (\%) & (\%) & (\%) &       & (\%)\\
\hline
1 A&$4.36\, 10^{-2}	$&$1.4	$&$2.5	$&$2.1	$&$1.3	$&$0.7	$&$0.5	$&$1.0	$&$0.95	$&$0.7$ \\ 
1 B&$3.37\, 10^{-2}	$&$1.6	$&$3.6	$&$2.9	$&$2.1	$&$0.4	$&$0.8	$&$1.9	$&$0.93	$&$1.7$ \\ 
1 C&$9.22\, 10^{-3}	$&$2.9	$&$6.1	$&$4.8	$&$3.8	$&$0.4	$&$1.6	$&$3.4	$&$0.92	$&$3.2$ \\ 
1 D&$1.88\, 10^{-3}	$&$6.5	$&$10.5	$&$8.8	$&$5.6	$&$1.2	$&$2.4	$&$4.9	$&$0.91	$&$1.3$ \\ 
\hline 
2 A&$4.20\, 10^{-2}	$&$1.7	$&$2.8	$&$2.5	$&$1.3	$&$0.9	$&$0.6	$&$0.7	$&$0.95	$&$0.6$ \\ 
2 B&$4.44\, 10^{-2}	$&$1.6	$&$2.8	$&$2.3	$&$1.7	$&$0.2	$&$0.6	$&$1.6	$&$0.93	$&$2.7$ \\ 
2 C&$1.26\, 10^{-2}	$&$2.8	$&$5.5	$&$4.4	$&$3.3	$&$0.7	$&$1.0	$&$3.1	$&$0.93	$&$1.3$ \\ 
2 D&$2.48\, 10^{-3}	$&$6.3	$&$10.6	$&$9.1	$&$5.4	$&$2.1	$&$1.0	$&$4.9	$&$0.91	$&$1.5$ \\ 
\hline 
3 A&$3.82\, 10^{-2}	$&$2.0	$&$3.1	$&$2.8	$&$1.2	$&$1.1	$&$0.3	$&$0.4	$&$0.93	$&$1.1$ \\ 
3 B&$5.86\, 10^{-2}	$&$1.5	$&$2.4	$&$2.1	$&$1.3	$&$0.2	$&$0.4	$&$1.2	$&$0.92	$&$1.2$ \\ 
3 B&$1.93\, 10^{-2}	$&$2.5	$&$4.9	$&$3.9	$&$2.9	$&$0.5	$&$0.6	$&$2.8	$&$0.93	$&$1.8$ \\ 
3 D&$3.78\, 10^{-3}	$&$5.6	$&$9.1	$&$7.7	$&$4.9	$&$1.0	$&$1.8	$&$4.5	$&$0.91	$&$2.9$ \\ 
\hline 
4 A&$2.36\, 10^{-2}	$&$2.9	$&$4.2	$&$4.0	$&$1.4	$&$1.3	$&$0.1	$&$0.3	$&$0.92	$&$1.2$ \\ 
4 B&$7.22\, 10^{-2}	$&$1.6	$&$2.5	$&$2.2	$&$1.2	$&$0.5	$&$0.4	$&$1.1	$&$0.91	$&$1.7$ \\ 
4 C&$3.91\, 10^{-2}	$&$2.1	$&$3.6	$&$3.0	$&$2.0	$&$0.5	$&$0.4	$&$1.9	$&$0.91	$&$0.9$ \\ 
4 D&$7.01\, 10^{-3}	$&$4.8	$&$8.0	$&$6.9	$&$4.2	$&$1.2	$&$1.0	$&$3.9	$&$0.93	$&$3.4$ \\ 
\hline 
5 A&$2.91\, 10^{-3}	$&$8.7	$&$8.9	$&$8.8	$&$1.1	$&$0.6	$&$0.8	$&$0.5	$&$0.92	$&$3.2$ \\ 
5 B&$4.50\, 10^{-2}	$&$2.3	$&$2.9	$&$2.7	$&$0.9	$&$0.8	$&$0.3	$&$0.5	$&$0.91	$&$3.2$ \\ 
5 C&$8.09\, 10^{-2}	$&$1.7	$&$2.5	$&$2.2	$&$1.1	$&$0.1	$&$0.2	$&$1.1	$&$0.91	$&$2.6$ \\ 
5 D&$4.54\, 10^{-2}	$&$2.1	$&$4.7	$&$4.2	$&$2.2	$&$0.7	$&$0.7	$&$1.9	$&$0.90	$&$2.9$ \\ 
\hline 
6 A&---	&---	&---	&---	&---	&---	&---	&---	&--- 	&---\\ 
6 B&---	&---	&---	&---	&---	&---	&---	&---	&--- 	&---\\ 
6 C&---	&---	&---	&---	&---	&---	&---	&---	&--- 	&---\\ 
6 D&$1.80\, 10^{-1}	$&$4.6	$&$7.6	$&$7.5	$&$1.5	$&$1.0	$&$0.3	$&$1.1	$&$0.91	$&$3.3$ \\ 
\hline 
\end{tabular} 
\normalfont
 \vspace{\baselineskip} 
\caption{\label{tab::2jet_Q2Ksi} Normalised 2-jet cross sections as a function of $Q^2$ and $\xi$ together with their relative errors and hadronisation correction factors. Other details are given in the caption to table \ref{tab::jet_Q2}. The bin nomenclature is defined in table \ref{tBinning}. At high $Q^2$ small $\xi$ values are kinematically disfavoured or forbidden.} 
\end{table} 
 
\newpage 
 
\begin{table}[p]
	  
\centering 

\renewcommand{\arraystretch}{1.10}	  
\begin{tabular}{| c | c || r | r | r | r || r | r | r | r || r |} \hline

\multicolumn{2}{|c||}{\multirow{2}{*}{\footnotesize{$150 < Q^2 < 200~\gevsq$} }}         
																 & \multicolumn{4}{|c||}{jet} &\multicolumn{4}{|c||}{2-jet} &      3-jet\\ \cline{3-11} 
\multicolumn{2}{|c||}{}		      & 1 a   &1 b    &1 c    &1 d    &1 a    &1 b    &1 c    &1 d    &1      \\ \hline \hline
\multirow{4}{*}{jet}    & 1 a   & $100  $ &$16  $ &$5   $ &$1   $ &$59  $ &$19  $ &$2   $ &$0   $ &$26  $\\ \cline{2-11} 
                        & 1 b   & $16   $ &$100 $ &$12  $ &$2   $ &$22  $ &$72  $ &$12  $ &$1   $ &$30  $\\ \cline{2-11} 
                        & 1 c   & $5    $ &$12  $ &$100 $ &$8   $ &$0   $ &$19  $ &$77  $ &$6   $ &$19  $\\ \cline{2-11} 
                        & 1 d   & $1    $ &$2   $ &$8   $ &$100 $ &$0   $ &$0   $ &$16  $ &$78  $ &$6   $\\ \hline \hline 
\multirow{4}{*}{2-jet}  & 1 a   & $59   $ &$22  $ &$0   $ &$0   $ &$100 $ &$0   $ &$0   $ &$0   $ &$21  $\\ \cline{2-11} 
                        & 1 b   & $19   $ &$72  $ &$19  $ &$0   $ &$0   $ &$100 $ &$0   $ &$0   $ &$30  $\\ \cline{2-11} 
                        & 1 c   & $2    $ &$12  $ &$77  $ &$16  $ &$0   $ &$0   $ &$100 $ &$0   $ &$16  $\\ \cline{2-11} 
                        & 1 d   & $0    $ &$1   $ &$6   $ &$78  $ &$0   $ &$0   $ &$0   $ &$100 $ &$4   $\\ \hline 
3-jet                   & 1     & $26   $ &$30  $ &$19  $ &$6   $ &$21  $ &$30  $ &$16  $ &$4   $ &$100 $\\ \hline 
\multicolumn{11}{c}{ }\\

\hline
\multicolumn{2}{|c||}{\multirow{2}{*}{\footnotesize{$200 < Q^2 < 270~\gevsq$}}}         
																& \multicolumn{4}{|c||}{jet} &\multicolumn{4}{|c||}{2-jet} &      3-jet\\ \cline{3-11} 
\multicolumn{2}{|c||}{}  				& 2 a   &2 b    &2 c    &2 d    &2 a    &2 b    &2 c    &2 d    &2      \\ \hline \hline
\multirow{4}{*}{jet}    & 2 a   & $100  $ &$16  $ &$4   $ &$1   $ &$58  $ &$19  $ &$2   $ &$1   $ &$25  $\\ \cline{2-11} 
                        & 2 b   & $16   $ &$100 $ &$13  $ &$2   $ &$22  $ &$71  $ &$13  $ &$1   $ &$29  $\\ \cline{2-11} 
                        & 2 c   & $4    $ &$13  $ &$100 $ &$9   $ &$0   $ &$20  $ &$76  $ &$8   $ &$20  $\\ \cline{2-11} 
                        & 2 d   & $1    $ &$2   $ &$9   $ &$100 $ &$0   $ &$0   $ &$14  $ &$75  $ &$8   $\\ \hline \hline 
\multirow{4}{*}{2-jet}  & 2 a   & $58   $ &$22  $ &$0   $ &$0   $ &$100 $ &$0   $ &$0   $ &$0   $ &$21  $\\ \cline{2-11} 
                        & 2 b   & $19   $ &$71  $ &$20  $ &$0   $ &$0   $ &$100 $ &$0   $ &$0   $ &$28  $\\ \cline{2-11} 
                        & 2 c   & $2    $ &$13  $ &$76  $ &$14  $ &$0   $ &$0   $ &$100 $ &$0   $ &$17  $\\ \cline{2-11} 
                        & 2 d   & $1    $ &$1   $ &$8   $ &$75  $ &$0   $ &$0   $ &$0   $ &$100 $ &$6   $\\ \hline \hline 
3-jet                   & 2     & $25   $ &$29  $ &$20  $ &$8   $ &$21  $ &$28  $ &$17  $ &$6   $ &$100 $\\ \hline 
\multicolumn{11}{c}{ }\\

\hline
\multicolumn{2}{|c||}{\multirow{2}{*}{\footnotesize{$270 < Q^2 < 400~\gevsq$}}}         
																& \multicolumn{4}{|c||}{jet} &\multicolumn{4}{|c||}{2-jet} &      3-jet\\ \cline{3-11} 
\multicolumn{2}{|c||}{}  				& 3 a   &3 b    &3 c    &3 d    &3 a    &3 b    &3 c    &3 d    &3      \\ \hline \hline
\multirow{4}{*}{jet}    & 3 a   & $100  $ &$16  $ &$5   $ &$1   $ &$59  $ &$19  $ &$3   $ &$0   $ &$27  $\\ \cline{2-11} 
                        & 3 b   & $16   $ &$100 $ &$13  $ &$2   $ &$20  $ &$71  $ &$12  $ &$1   $ &$30  $\\ \cline{2-11} 
                        & 3 c   & $5    $ &$13  $ &$100 $ &$8   $ &$0   $ &$20  $ &$77  $ &$7   $ &$21  $\\ \cline{2-11} 
                        & 3 d   & $1    $ &$2   $ &$8   $ &$100 $ &$0   $ &$0   $ &$16  $ &$77  $ &$8   $\\ \hline \hline 
\multirow{4}{*}{2-jet}  & 3 a   & $59   $ &$20  $ &$0   $ &$0   $ &$100 $ &$0   $ &$0   $ &$0   $ &$21  $\\ \cline{2-11} 
                        & 3 b   & $19   $ &$71  $ &$20  $ &$0   $ &$0   $ &$100 $ &$0   $ &$0   $ &$30  $\\ \cline{2-11} 
                        & 3 c   & $3    $ &$12  $ &$77  $ &$16  $ &$0   $ &$0   $ &$100 $ &$0   $ &$20  $\\ \cline{2-11} 
                        & 3 d   & $0    $ &$1   $ &$7   $ &$77  $ &$0   $ &$0   $ &$0   $ &$100 $ &$6   $\\ \hline \hline 
3-jet                   & 3     & $27   $ &$30  $ &$21  $ &$8   $ &$21  $ &$30  $ &$20  $ &$6   $ &$100 $\\ \hline 

\end{tabular}

 \vspace{\baselineskip} 
    	\caption{\label{tMatrix13} The statistical correlation factors 
	  given in percent between different $P_{T,\text{obs}}$ bins 
	  of different jet observables inside $Q^2$ bins $1$ to $3$ 
	  as estimated from the data. The bin nomenclature is defined 
	  in table \ref{tBinning}.}

\end{table}

\begin{table}[b]

\centering 
	\renewcommand{\arraystretch}{1.10} 
\begin{tabular}{| c | c || r | r | r | r || r | r | r | r || r |} \hline

\multicolumn{2}{|c||}{\multirow{2}{*}{\footnotesize{$400 < Q^2 < 700~\gevsq$}}}         
																& \multicolumn{4}{|c||}{jet} &\multicolumn{4}{|c||}{2-jet} &      3-jet\\ \cline{3-11} 
\multicolumn{2}{|c||}{}  & 4 a   &4 b    &4 c    &4 d    &4 a    &4 b    &4 c    &4 d    &4      \\ \hline \hline
\multirow{4}{*}{jet}    & 4 a   & $100  $ &$15  $ &$6   $ &$1   $ &$58  $ &$20  $ &$3   $ &$1   $ &$28  $\\ \cline{2-11} 
                        & 4 b   & $15   $ &$100 $ &$13  $ &$2   $ &$19  $ &$70  $ &$14  $ &$1   $ &$31  $\\ \cline{2-11} 
                        & 4 c   & $6    $ &$13  $ &$100 $ &$9   $ &$0   $ &$21  $ &$76  $ &$7   $ &$23  $\\ \cline{2-11} 
                        & 4 d   & $1    $ &$2   $ &$9   $ &$100 $ &$0   $ &$0   $ &$15  $ &$78  $ &$8   $\\ \hline \hline 
\multirow{4}{*}{2-jet}  & 4 a   & $58   $ &$19  $ &$0   $ &$0   $ &$100 $ &$0   $ &$0   $ &$0   $ &$23  $\\ \cline{2-11} 
                        & 4 b   & $20   $ &$70  $ &$21  $ &$0   $ &$0   $ &$100 $ &$0   $ &$0   $ &$30  $\\ \cline{2-11} 
                        & 4 c   & $3    $ &$14  $ &$76  $ &$15  $ &$0   $ &$0   $ &$100 $ &$0   $ &$20  $\\ \cline{2-11} 
                        & 4 d   & $1    $ &$1   $ &$7   $ &$78  $ &$0   $ &$0   $ &$0   $ &$100 $ &$7   $\\ \hline \hline 
3-jet                   & 4     & $28   $ &$31  $ &$23  $ &$8   $ &$23  $ &$30  $ &$20  $ &$7   $ &$100 $\\ \hline 
\multicolumn{11}{c}{ }\\

\hline
\multicolumn{2}{|c||}{\multirow{2}{*}{\footnotesize{$700 < Q^2 < 5000~\gevsq$}}}         
																& \multicolumn{4}{|c||}{jet} &\multicolumn{4}{|c||}{2-jet} &      3-jet\\ \cline{3-11} 
\multicolumn{2}{|c||}{}  & 5 a   &5 b    &5 c    &5 d    &5 a    &5 b    &5 c    &5 d    &5      \\ \hline \hline
\multirow{4}{*}{jet}    & 5 a   & $100  $ &$16  $ &$5   $ &$3   $ &$57  $ &$21  $ &$5   $ &$1   $ &$28  $\\ \cline{2-11} 
                        & 5 b   & $16   $ &$100 $ &$13  $ &$2   $ &$20  $ &$68  $ &$14  $ &$1   $ &$30  $\\ \cline{2-11} 
                        & 5 c   & $5    $ &$13  $ &$100 $ &$9   $ &$0   $ &$20  $ &$71  $ &$9   $ &$21  $\\ \cline{2-11} 
                        & 5 d   & $3    $ &$2   $ &$9   $ &$100 $ &$0   $ &$0   $ &$23  $ &$72  $ &$7   $\\ \hline \hline 
\multirow{4}{*}{2-jet}  & 5 a   & $57   $ &$20  $ &$0   $ &$0   $ &$100 $ &$0   $ &$0   $ &$0   $ &$18  $\\ \cline{2-11} 
                        & 5 b   & $21   $ &$68  $ &$20  $ &$0   $ &$0   $ &$100 $ &$0   $ &$0   $ &$29  $\\ \cline{2-11} 
                        & 5 c   & $5    $ &$14  $ &$71  $ &$23  $ &$0   $ &$0   $ &$100 $ &$0   $ &$19  $\\ \cline{2-11} 
                        & 5 d   & $1    $ &$1   $ &$9   $ &$72  $ &$0   $ &$0   $ &$0   $ &$100 $ &$7   $\\ \hline \hline 
3-jet                   & 5     & $28   $ &$30  $ &$21  $ &$7   $ &$18  $ &$29  $ &$19  $ &$7   $ &$100 $\\ \hline 
\multicolumn{11}{c}{ }\\

\hline
\multicolumn{2}{|c||}{\multirow{2}{*}{\footnotesize{$5000 < Q^2 < 15000~\gevsq$}}}         
																& \multicolumn{4}{|c||}{jet} &\multicolumn{4}{|c||}{2-jet} &      3-jet\\ \cline{3-11} 
\multicolumn{2}{|c||}{}  & 6 a   &6 b    &6 c    &6 d    &6 a    &6 b    &6 c    &6 d    &6      \\ \hline \hline
\multirow{4}{*}{jet}    & 6 a   & $100  $ &$16  $ &$6   $ &$3   $ &$53  $ &$22  $ &$7   $ &$0   $ &$28  $\\ \cline{2-11} 
                        & 6 b   & $16   $ &$100 $ &$12  $ &$3   $ &$20  $ &$62  $ &$15  $ &$2   $ &$32  $\\ \cline{2-11} 
                        & 6 c   & $6    $ &$12  $ &$100 $ &$7   $ &$0   $ &$21  $ &$58  $ &$9   $ &$22  $\\ \cline{2-11} 
                        & 6 d   & $3    $ &$3   $ &$7   $ &$100 $ &$0   $ &$0   $ &$22  $ &$67  $ &$14  $\\ \hline \hline 
\multirow{4}{*}{2-jet}  & 6 a   & $53   $ &$20  $ &$0   $ &$0   $ &$100 $ &$0   $ &$0   $ &$0   $ &$19  $\\ \cline{2-11} 
                        & 6 b   & $22   $ &$62  $ &$21  $ &$0   $ &$0   $ &$100 $ &$0   $ &$0   $ &$28  $\\ \cline{2-11} 
                        & 6 c   & $7    $ &$15  $ &$58  $ &$22  $ &$0   $ &$0   $ &$100 $ &$0   $ &$25  $\\ \cline{2-11} 
                        & 6 d   & $0    $ &$2   $ &$9   $ &$67  $ &$0   $ &$0   $ &$0   $ &$100 $ &$13  $\\ \hline \hline 
3-jet                   & 6     & $28   $ &$32  $ &$22  $ &$14  $ &$19  $ &$28  $ &$25  $ &$13  $ &$100 $\\ \hline 

\end{tabular}

  \vspace{\baselineskip} 
     	\caption{\label{tMatrix46} The statistical correlation factors
	  given in percent between different $P_{T,\text{obs}}$ bins 
	  of different jet observables inside $Q^2$ bins $4$ to $6$ as 
	  estimated from the data. The bin nomenclature is defined in 
	  table \ref{tBinning}.}

\end{table}

\begin{table}[p]
	\centering
	\renewcommand{\arraystretch}{1.80}
  \begin{tabular}{| l | c | c | c| c | c |}
			\hline 
     	\multirow{3}{*}{Measurement} & 	\multirow{3}{*}{$\alpha_S(M_Z)$} & \multicolumn{3}{|c|}{Uncertainty} &	\multirow{3}{*}{$\chi^2/\rm ndf$} \\ \cline{3-5} 

     	& & experimental & theory  & {\sc pdf} & \\

		\hline        
     $\frac{\sigma_{\rm jet}}{\sigma_{\rm NC}}\left(Q^2,P_T\right)$ &
     0.1195 & 0.0010 & $^{+0.0049}_{-0.0036}$ & 0.0018 & 24.7/23 \\ \hline
    $\frac{\sigma_{\textnormal{2-jet}}}{\sigma_{\rm NC}}\left(Q^2,\left\langle P_T \right\rangle \right)$ &
     0.1155 & 0.0009 & $^{+0.0042}_{-0.0031}$ & 0.0017 & 30.4/23 \\ \hline
  $\frac{\sigma_{\textnormal{3-jet}}}{\sigma_{\rm NC}}\left(Q^2 \right)$ &
     0.1172 & 0.0013 & $^{+0.0052}_{-0.0031}$ & 0.0009 & 7.0/5 \\ \hline  \hline 
     $\frac{\sigma_{\rm jet}}{\sigma_{\rm NC}},~\frac{\sigma_{\textnormal{2-jet}}}{\sigma_{\rm NC}},~\frac{\sigma_{\textnormal{3-jet}}}{\sigma_{\rm NC}}$ &
     0.1168 & 0.0007 & $^{+0.0046}_{-0.0030}$ & 0.0016 & 65.0/53 \\ \hline    
  \end{tabular}
  \vspace{\baselineskip} 
	\caption{\label{tab::Fits} Values of $\alpha_s (M_Z)$ obtained from fits to the individual normalised inclusive jet, 2-jet and 3-jet cross sections and from a simultaneous fit to all of them. Fitted values are given with experimental, theoretical and PDF errors as well as with the normalised $\chi^2$ of the fit.}
\end{table}

\begin{thebibliography}{99}


\bibitem{BreitFrame} 
R. P. Feynman, "Photon-Hadron Interactions", Benjamin, New York (1972).

\bibitem{Adloff:2000tq}
  C.~Adloff {\it et al.}  [H1 Collaboration],
  Eur.\ Phys.\ J.\  C {\bf 19} (2001) 289 [hep-ex/0010054].


\bibitem{H1Incl:2007pb}
  A.~Aktas {\it et al.}  [H1 Collaboration],
  Phys.\ Lett.\ B {\bf 653} (2007) 134 [arXiv:0706.3722].

\bibitem{Chekanov:2006yc}
  S.~Chekanov {\it et al.}  [ZEUS Collaboration],
  Phys.\ Lett.\  B {\bf 649} (2007) 12
  [hep-ex/0701039].

  
\bibitem{Abt:1996hi}
  I.~Abt {\it et al.}
  [H1 Collaboration],
  Nucl.\ Instrum.\ Meth.\ A {\bf 386} (1997) 310,
  ibid., 348.

\bibitem{Appuhn:1996na}
  R.~D.~Appuhn {\it et al.}  [H1 SPACAL Group],
  Nucl.\ Instrum.\ Meth.\ A {\bf 386} (1997) 397.


\bibitem{h1cal}
B.~Andrieu {\it et al.}  [H1 Calorimeter Group],
Nucl.\ Instrum.\ Meth.\ A {\bf 336} (1993) 460.
%

\bibitem{Andrieu:1993tz}
  B.~Andrieu {\it et al.}  [H1 Calorimeter Group],
  Nucl.\ Instrum.\ Meth.\ A {\bf 336} (1993) 499;

  B.~Andrieu {\it et al.}  [H1 Calorimeter Group],
  Nucl.\ Instrum.\ Meth.\ A {\bf 350} (1994) 57.

\bibitem{Pitzl:2000wz}
  D.~Pitzl {\it et al.},
  Nucl.\ Instrum.\ Meth.\  A {\bf 454} (2000) 334
[hep-ex/0002044];
  
   B.~List {\it et al.}, 
   Nucl.\ Instrum.\ Meth.\  A {\bf 549} (2005) 33.


\bibitem{Adloff:2003uh}
  C.~Adloff {\it et al.}  [H1 Collaboration],
  Eur.\ Phys.\ J.\  C {\bf 30} (2003) 1
  [hep-ex/0304003].


\bibitem{Peez:2003zd}
  M.~Peez,
  "Search for deviations from the standard model in high transverse energy
  processes at the electron proton collider HERA" (in French), CPPM-T-2003-04 (available at http://www-h1.desy.de/psfiles/theses/).

\bibitem{Portheault:2005uu}
  B.~Portheault,
  "First measurement of charged and neutral current cross sections
  with the polarized positron beam at HERA II and QCD-electroweak
  analyses" (in French), LAL-05-05 (available at 
  http://www-h1.desy.de/psfiles/theses/).

  

\bibitem{Bassler:1994uq}
  U.~Bassler and G.~Bernardi,
  Nucl.\ Instrum.\ Meth.\ A {\bf 361} (1995) 197 [hep-ex/9412004];

  U.~Bassler and G.~Bernardi,
  Nucl.\ Instrum.\ Meth.\ A {\bf 426} (1999) 583 [hep-ex/9801017].

\bibitem{Ellis:1993tq}
  S.~D.~Ellis and D.~E.~Soper,
  Phys.\ Rev.\ D {\bf 48} (1993) 3160
[hep-ph/9305266];
  
  S.~Catani {\it et al.},
  Nucl.\ Phys.\ B {\bf 406} (1993) 187.
  
\bibitem{Frixione:1997ks}
  S.~Frixione and G.~Ridolfi,
  Nucl.\ Phys.\  B {\bf 507} (1997) 315
  [hep-ph/9707345].



\bibitem{Charchula:1994kf}
  K.~Charchula, G.~A.~Schuler and H.~Spiesberger,
  DJANGOH 1.4,
  Comput.\ Phys.\ Commun.\  {\bf 81} (1994) 381.

\bibitem{Lonnblad:1992tz}
  L.~L\"onnblad,
  ARIADNE 4.08,
  Comput.\ Phys.\ Commun.\  {\bf 71} (1992) 15.

\bibitem{Jung:1993gf}
  H.~Jung,
 RAPGAP 2.08,
  Comput.\ Phys.\ Commun.\  {\bf 86} (1995) 147.

\bibitem{Andersson:1983ia}
  B.~Andersson, G.~Gustafson, G.~Ingelman and T.~Sjostrand, JETSET 7.41,
  Phys.\ Rept.\  {\bf 97} (1983) 31.

\bibitem{Brun:1987ma}
  R.~Brun {\it et al.}, "GEANT3", CERN-DD/EE/84-1.
  
\bibitem{Gouzevitch:2008zz}
  M.~Gouzevitch,
  "Measurement of the strong coupling constant alpha(s) with hadronic
  jets in Deep Inelastic Scattering" (in French), DESY-THESIS-2008-047
  (available at http://www-h1.desy.de/psfiles/theses/).

\bibitem{Kwiatkowski:1990es}
  A.~Kwiatkowski, H.~Spiesberger and H.~J.~M\"ohring, HERACLES 4.63,
  Comput.\ Phys.\ Commun.\  {\bf 69} (1992) 155.

\bibitem{Ingelman:1996mq}
  G.~Ingelman, A.~Edin and J.~Rathsman,
  LEPTO 6.5,
  Comput.\ Phys.\ Commun.\  {\bf 101} (1997) 108
[hep-ph/9605286].

\bibitem{BIB::FACTOR_JET}
B.R.~Webber, 
J. Phys. \textbf{G19} (1993) 1567.


\bibitem{Nagy:2001xb}
  Z.~Nagy and Z.~Trocsanyi, NLOJET++ 4.0.1,
  Phys.\ Rev.\ Lett.\  {\bf 87} (2001) 082001
[hep-ph/0104315].

\bibitem{Catani:1996vz}
  S.~Catani and M.~H.~Seymour,
  Nucl.\ Phys.\  B {\bf 485} (1997) 291
  [Erratum-ibid.\  B {\bf 510} (1998) 503]
[hep-ph/9605323].

\bibitem{Kluge:2006xs}
  T.~Kluge, K.~Rabbertz and M.~Wobisch, FastNLO 1.0,
  hep-ph/0609285, 
published in "Deep inelastic scattering DIS2006, Proceedings of the 14th
workshop", eds. M.~Kuze, K.~Nagano, K.~Tokushuka, 483.
    
\bibitem{Tung:2006tb}
  W.~K.~Tung {\it et al.}
  JHEP {\bf 0702} (2007) 053
[hep-ph/0611254].

\bibitem{Dasgupta:2007wa}

  M.~Dasgupta and Y.~Delenda,
  JHEP {\bf 0907} (2009) 004
  [arXiv:0903.2187];



  M.~Dasgupta, L.~Magnea and G.~P.~Salam,
  JHEP {\bf 0802} (2008) 055
  [arXiv:0712.3014];
      
  M.~Dasgupta and B.~R.~Webber,
  Eur.\ Phys.\ J.\  C {\bf 1} (1998) 539
  [hep-ph/9704297];

  M.~Dasgupta and B.~R.~Webber,
  JHEP {\bf 9810} (1998) 001
  [hep-ph/9809247].
  
\bibitem{Dokshitzer:1995qm}
  Y.~L.~Dokshitzer, G.~Marchesini and B.~R.~Webber,
  Nucl.\ Phys.\  B {\bf 469} (1996) 93
  [hep-ph/9512336].

\bibitem{Aktas:2005tz}
  A.~Aktas {\it et al.}  [H1 Collaboration],
  Eur.\ Phys.\ J.\  C {\bf 46} (2006) 343
  [hep-ex/0512014].
  


\bibitem{bib:chifit}
C. Pascaud and F. Zomer, preprint LAL 95-05 (1995).

\bibitem{Botje:1999dj}
  M.~Botje,
  Eur.\ Phys.\ J.\ C {\bf 14} (2000) 285
[hep-ph/9912439].

\bibitem{Barone:1999yv}
  V.~Barone, C.~Pascaud and F.~Zomer,
  Eur.\ Phys.\ J.\ C {\bf 12} (2000) 243
[hep-ph/9907512].

\bibitem{Adloff:2000dp}
  C.~Adloff {\it et al.} [H1 Collaboration],
  Phys.\ Lett.\ B {\bf 479} (2000) 358
[hep-ex/0003002].


\bibitem{Jones:2003yv}
  R.~W.~L.~Jones {\em et al.},
  JHEP {\bf 0312} (2003) 007
  [hep-ph/0312016].
  
\bibitem{Martin:2009iq}
  A.~D.~Martin, W.~J.~Stirling, R.~S.~Thorne and G.~Watt,
  arXiv:0901.0002.

\bibitem{Cacciari:2008gp}
  M.~Cacciari, G.~P.~Salam and G.~Soyez,
  JHEP {\bf 0804} (2008) 063 [arXiv:0802.1189].

\bibitem{Amsler:2008zzb}
  C.~Amsler {\it et al.}  [Particle Data Group],
  Phys.\ Lett.\  B {\bf 667} (2008) 1.
  
\bibitem{Bethke:2006ac}
  S.~Bethke,
  Prog.\ Part.\ Nucl.\ Phys.\  {\bf 58} (2007) 351 
  [hep-ex/0606035].

\bibitem{Chekanov:2005ve}
  S.~Chekanov {\it et al.}  [ZEUS Collaboration],
  Eur.\ Phys.\ J.\  C {\bf 44} (2005) 183
  [hep-ex/0502007].

\bibitem{Dissertori:2007xa}
  G.~Dissertori {\it et al.},
  JHEP {\bf 0802} (2008) 040
  [arXiv:0712.0327].


\end{thebibliography}
\end{document}